\documentclass[twocolumn]{aastex63}

\usepackage{booktabs}

\received{}
\revised{}
\accepted{}
\submitjournal{ApJ}

\newcommand{\C}[1]{{\color{cyan} \bf #1}}
\newcommand{\R}[1]{{\color{red} \bf #1}}
\newcommand{\N}[1]{{\color{black} \bf #1}}

\newcommand{\Msun}{${\rm M}_{\odot}$}

\newcommand{\SiII}{Si~{\sc ii}}
\newcommand{\SII}{S~{\sc ii}}

\newcommand{\NaI}{Na~{\sc i}~D}

\newcommand{\FeII}{Fe~{\sc ii}}
\newcommand{\CaII}{Ca~{\sc ii}}

\newcommand{\Nifs}{$^{56}$Ni}
\newcommand{\ld}{$\lambda$}
\shorttitle{SN 2021fxy with detached \SiII~HVFs}
\shortauthors{Li et al.}

\begin{document}
	
\title{Optical observations on the young Type Ia SN 2021fxy with detached high velocity features}

\author[0009-0003-3758-0598]{Liping Li}
\affiliation{International Centre of Supernovae (ICESUN), Yunnan Key Laboratory of Supernova Research, Yunnan Observatories, Chinese Academy of Sciences (CAS), Kunming, 650216, China}

\author[0000-0002-8296-2590]{Jujia Zhang}
\affiliation{International Centre of Supernovae (ICESUN), Yunnan Key Laboratory of Supernova Research, Yunnan Observatories, Chinese Academy of Sciences (CAS), Kunming, 650216, China}
\email{jujia@ynao.ac.cn}	

\author[0009-0009-2664-8212]{Zhenyu Wang} 
\affiliation{International Centre of Supernovae (ICESUN), Yunnan Key Laboratory of Supernova Research, Yunnan Observatories, Chinese Academy of Sciences (CAS), Kunming, 650216, China}
\affiliation{School of Astronomy and Space Science, University of Chinese Academy of Sciences, Beijing 100049,1408, People's Republic of China}

\author[0000-0002-7334-2357]{Xiaofeng Wang}
\affiliation{Physics Department and Tsinghua Center for Astrophysics (THCA), Tsinghua University, Beijing, 100084, China}	
	
\author[0009-0002-3956-6143]{Qian Zhai}
\affiliation{International Centre of Supernovae (ICESUN), Yunnan Key Laboratory of Supernova Research, Yunnan Observatories, Chinese Academy of Sciences (CAS), Kunming, 650216, China}

\email{zhaiqian@ynao.ac.cn}

\author[0009-0004-4256-1209]{Shengyu Yan}
\affiliation{Physics Department and Tsinghua Center for Astrophysics (THCA), Tsinghua University, Beijing, 100084, China}

\author[0000-0002-3231-1167]{Bo Wang} 
\affiliation{International Centre of Supernovae (ICESUN), Yunnan Key Laboratory of Supernova Research, Yunnan Observatories, Chinese Academy of Sciences (CAS), Kunming, 650216, China}

\author{Jinming Bai}
\affiliation{International Centre of Supernovae (ICESUN), Yunnan Key Laboratory of Supernova Research, Yunnan Observatories, Chinese Academy of Sciences (CAS), Kunming, 650216, China}

	
	
\begin{abstract}
We present optical observations on the young type Ia supernova (SN Ia) SN 2021fxy obtained within a few days after the explosion, with a focus on its prominent high-velocity features (HVFs). 
It reached a \textit{B}-band maximum of $M_{\rm max}(B) = -19.36\pm0.31$ mag, corresponding to a bolometric luminosity of $\sim 1.3\times10^{43}~\rm{erg~s^{-1}}$ with a synthesized  \Nifs\,mass of $0.58\pm0.14$ \Msun. 
The early spectra exhibit strong HVFs of intermediate-mass elements that are significantly detached from the photospheric components. 
In particular, the velocity of the Si~II $\lambda6355$ HVFs follows a power-law evolution ($\beta \approx 0.1$), shallower than the expected photospheric velocity evolution expected for an assumed $n=10$ density profile ($\beta \approx 0.22$) under homologous expansion.
This behavior is consistent with the HVFs forming in intrinsic ejecta structures at least partially decoupled from the bulk outer ejecta, providing a possible constraint on the explosion physics of SNe Ia.
\end{abstract}
	
	
	
\section{Introduction} 
Type Ia supernovae (SNe Ia) are generally believed to result from thermonuclear explosions of carbon-oxygen white dwarfs (WDs) in binary systems as they approach the Chandrasekhar mass ($\sim1.4$ \Msun) via accretion or merger processes \citep{2000ARA&A..38..191H,2014ARA&A..52..107M}.
Due to the consistent energy supply mechanism from $^{56}$Ni decay, the optical light curves of SNe Ia exhibit highly consistent characteristics.
As a result, their peak luminosities can be standardized using the width–luminosity relation (WLR; \citealt{1993ApJ...413L.105P}).
Therefore, SNe Ia are excellent distance indicators on cosmic scales and provide the first evidence for accelerating expansion of the universe and reveal the existence of dark energy \citep{1998AJ....116.1009R,1999ApJ...517..565P}. 

he empirical standardization of SNe Ia is implemented through light curve fitters that quantify the WLR and the accompanying color–luminosity relation \citep{1996ApJ...473...88R,1998A&A...331..815T} in a data-driven way. The decline rate parameter $\Delta m_{15}$, defined as the drop in $B$-band magnitude within 15 days after peak brightness \citep{1993ApJ...413L.105P}, is the most-used metric to characterize light curve width.
It can be easily obtained by fitting polynomial to the light curves.
More recent analyses employ the widely-used \texttt{SALT2} \citep[the spectral adaptive light curve template;][]{2007A&A...466...11G} fitter, which describes each SN Ia by a stretch parameter $x_1$ (encoding the light curve width) and a colour parameter $c$ (more negative values correspond to bluer events). 
Beyond distance measurement, these empirical parameters serve as powerful diagnostics for probing the diversity of SN Ia physics and/or their environments.
	
	
The spectroscopic analysis reveals the diversity among normal SNe Ia, enabling the subdivision of these objects into distinct groups based on their spectral characteristics.
Based on the temporal velocity gradient of \SiII~\ld6355, \citet{2005ApJ...623.1011B} suggested that normal SNe Ia can be classified into three classes: FAINT SNe Ia, high velocity gradient (HVG) SNe Ia, and low velocity gradient (LVG) SNe Ia. 
Considering the expansion velocity of \SiII~\ld6355 near the $ B $-band maximum, SNe Ia can also be divided into two subclasses: normal velocity (NV) SNe Ia and high velocity (HV) SNe Ia \citep{2009ApJ...699L.139W}. 
Alternatively, based on the equivalent width (EW) of the absorption features of \SiII~\ld6355 and \ld5972 near the \textit{B}-band maximum, \citet{2006PASP..118..560B,2009PASP..121..238B} classified SNe Ia into four subclasses: cool (CL), shallow silicon (SS), core normal (CN), and broad line (BL).
Most recently, based on the relative EW of the absorption features of \SiII~\ld6355 near the \textit{B}-band maximum, \citet{2023ApJ...943..159M} suggested that SNe Ia can be divided into two subclasses: broad line (BL) and normal line (NL).
These subclasses reflect subtle differences in the explosions, including the nucleosynthesis, temperatures, ejecta dynamics, and are potentially linked to the original explosion mechanisms and environments.

In addition, high-velocity features (HVFs) for intermediate-mass elements (IMEs; e.g. \SiII, \CaII) are detectable in the early-phase spectra for some SNe Ia \citep[e.g.,][]{2005MNRAS.357..200M,2014MNRAS.437..338C,2015MNRAS.451.1973S,2016ApJ...817..114Z,2026ApJ...996...10L}, appearing as blueshifted absorption lines offset by 6,000-13,000 km s$^{-1}$ from the corresponding photospheric velocity features (PVFs).
Although their physical origin remains debated, HVFs are considered as the potential indicators of the explosion mechanisms or circumstellar material \citep[e.g.,][]{2005MNRAS.357..200M}. 
Early statistical studies show that HVFs of \CaII~ regularly appear in the early spectra, while HVFs of \SiII~are occurring in only $\sim30\%$ in SN Ia population \citep[][]{2015MNRAS.451.1973S}. 
However, \citet{2025A&A...695A.264H} recently demonstrated that the detection rates of \SiII~HVFs are significantly dependent on the phases of the spectra.
The \SiII~HVFs can be detected in most of the spectra at very early phases ($\sim75\%$ at  $<~-11$ days relative to their \textit{B}-band maximum).
It indicates that the \SiII~HVFs may be the common features in SN Ia population. 
Despite the diversity in HVFs among SNe Ia, the spectra observed at very early phases are important to reveal the nature of the detached HVFs.

%
%
SN 2021fxy is a SN Ia exhibiting prominent, detached \SiII~HVFs in the early-time spectra.
Its ultraviolet (UV) properties were analyzed in detail by \citet{2023MNRAS.522.3481D}, who proposed that the observed  UV flux suppression may arise from high-velocity ejecta.
In this paper, we present comprehensive optical photometric and spectroscopic observations of SN 2021fxy that trace its early evolution and fill the observational gaps identified by \citet{2023MNRAS.522.3481D}.
We focus on the evolution of the detached \SiII~HVFs at early phases and discuss the implications for the explosion physics.
	
This paper was organized as follows. Observations and data reductions of SN 2021fxy are described in Section \ref{sec:2}. Then we analyze its light/color curves in Section~\ref{sec:3} and the spectral evolution is presented in Section~\ref{sec:4}. In Section \ref{sec:5}, we discuss the high velocity features and the classification of SN 2021fxy. Finally, a conclusion is given in Section~\ref{sec:6}.	
	
\section{Observations and Data Reduction}\label{sec:2}
\begin{figure}
	\includegraphics[width=\columnwidth]{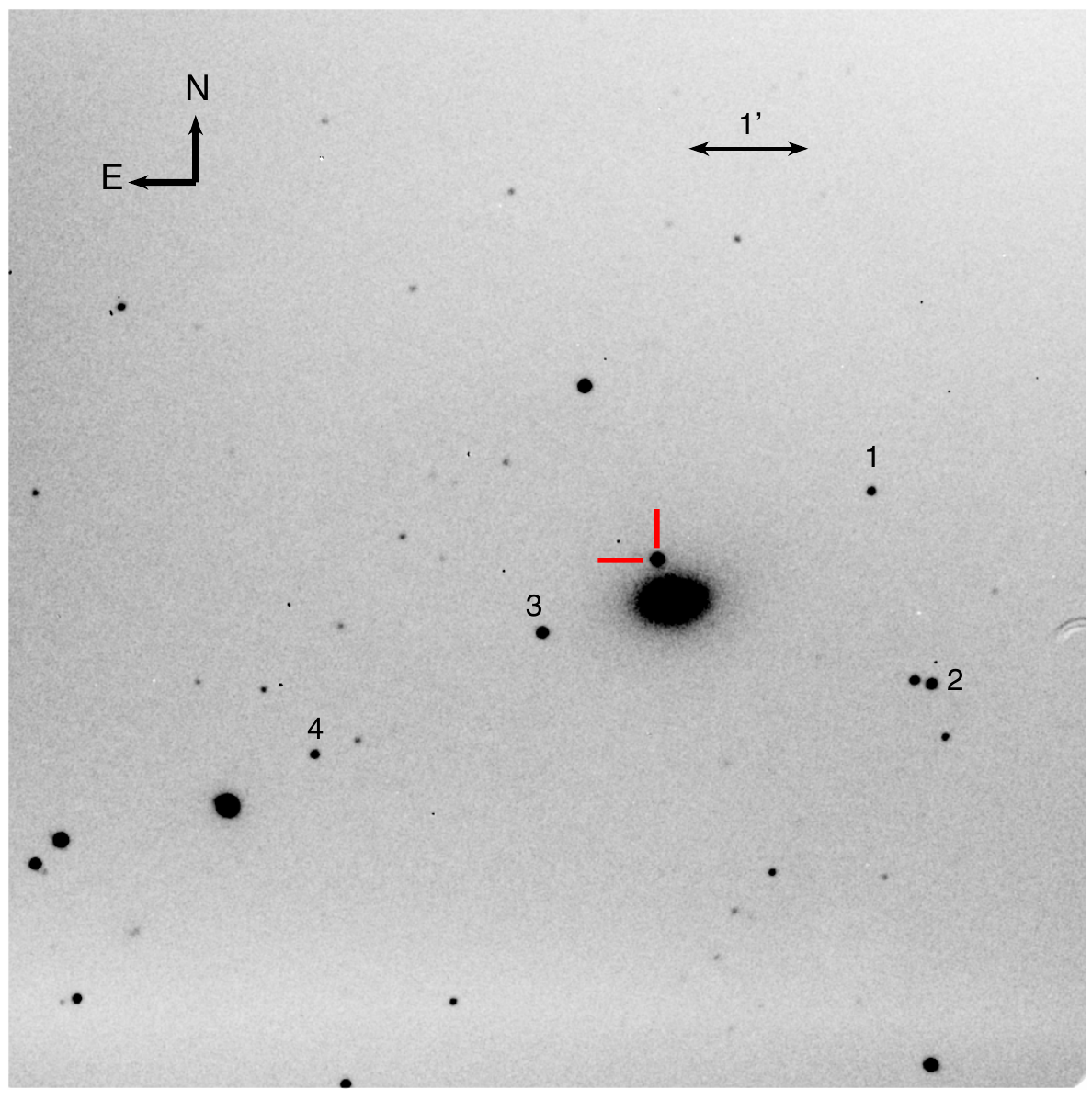}
	\caption{Finder chart of SN 2021fxy in NGC 5018, taken by the LJT. The supernova and four local reference stars are marked.  \label{fig:img}}
\end{figure}

SN 2021fxy was discovered by K. Itagaki on 2021 March 17.79, UT (universal time is used throughout this paper) with 16.9 mag \citep[clear filter;][]{2021TNSTR.785....1I}. It was then classified as a young SN Ia by Saurabh W. Jha \citep{2021TNSCR.813....1J}. 
Its coordinates are RA = {13$^h$}{13$^m$}{01$\fs$570}, DEC = -{19$\degr$}{30$\arcmin$}{45$\farcs$18} (J2000), located at 20.67$''$ north, 8.23$''$ east (3.2 Kpc) from the center of NGC 5018 \citep{1985AnTok..20..335T} (see Figure \ref{fig:img}). 
The redshift of the galaxy NGC 5018 is 0.0094.
This redshift is adopted in the analyses without any peculiar velocity correction for SN 2021fxy, while the Hubble parameter of $H_{0}=73~\rm{km~s^{-1}~Mpc^{-1}}$ is used throughout this paper.

Several hours after its discovery, we obtained a spectrum of SN 2021fxy with the Lijiang 2.4 m telescope (LJT; \citet{2015RAA....15..918F}) equipped with the Yunnan Faint Object Spectrograph and Camera (YFOSC; \citet{2019RAA....19..149W}), marking the start of our monitoring campaign.
Under the LiONS program (Lijiang One hour per Night for Supernova observation), we conducted a long-term photometric and spectroscopic campaign with LJT, spanning $t \approx  -$14 d to $t \approx  +$78 d relative to the $B$-band maximum ($ t $ is the time after $B$-band maximum and is used throughout this paper). 
Meanwhile, the Tsinghua-NAOC 0.8 meter telescope  \citep[TNT;][]{2008ApJ...675..626W, 2012RAA....12.1585H} at Xing-long Observation of National Astronomical Observatories (NAOC) contributed additional optical photometry. Furthermore, the Ultraviolet/Optical Telescope \citep[UVOT;][]{2005SSRv..120...95R} onboard the \textit{Swift} satellite was also used to collect the UV/optical data of SN 2021fxy at $t \approx $ -14 d.

\subsection{Photometry}
Broadband \textit{BV}- and Sloan \textit{gri}-band photometry of SN 2021fxy  obtained by the LJT and TNT covers the period from $t \approx -14$ d to  $t \approx +78$ d. 

The CCD images are reduced using the IRAF \footnote{IRAF,the Image Reduction and Analysis Facility, is distributed by the National Optical Astronomy Observatory, which is operated by the Association of Universities for Research in Astronomy (AURA), Inc. under cooperative agreement with the National Science Foundation (NSF).} standard procedure, including bias subtraction, flat fielding, and removal of cosmic rays. 
We perform the galaxy subtractions for all filters using template observations gathered in March 2022. The Saccadic Fast Fourier Transform (SFFT) algorithm \citep[][]{2022ApJ...936..157H} is used for the subtraction procedure. Aperture photometry is applied to the image after the template subtraction.
The instrumental magnitudes of SN 2021fxy are further converted to the standard Johnson \textit{BV} and Sloan \textit{gri} systems based on the 4 local standard stars (as labeled in Figure \ref{fig:img} and listed in Table \ref{tab:stand}). 
Note that these conversions do not include any correction by extinction or color terms.
The final results of the photometry from the LJT and TNT are listed in Table \ref{tab:Opti_Pho}.

The \textit{Swift}-UVOT observations cover a few days in two UV filters (\textit{uvw2} and \textit{uvw1}) and three broadband optical filters (\textit{U}, \textit{B} and \textit{V}). 
The UV-optical photometry data listed in Table \ref{tab:UVOT} are publicly available through the $Swift$ Optical/Ultraviolet Supernova Archive (SOUSA\footnote{\url{http://people.physics.tamu.edu/pbrown/SwiftSN/swift_sn.html}}; \citealp{2014Ap&SS.354...89B}).

\subsection{Spectroscopy}
All spectra obtained by LJT are reduced using standard IRAF routines. The spectra are calibrated with the spectrophotometric  standard stars observed at a similar air mass on the same night. Furthermore, the spectra are corrected for the continuum atmospheric extinction at the Lijiang Observatory.
The telluric lines in the spectra are also removed.

\begin{figure}
	\includegraphics[width=\columnwidth]{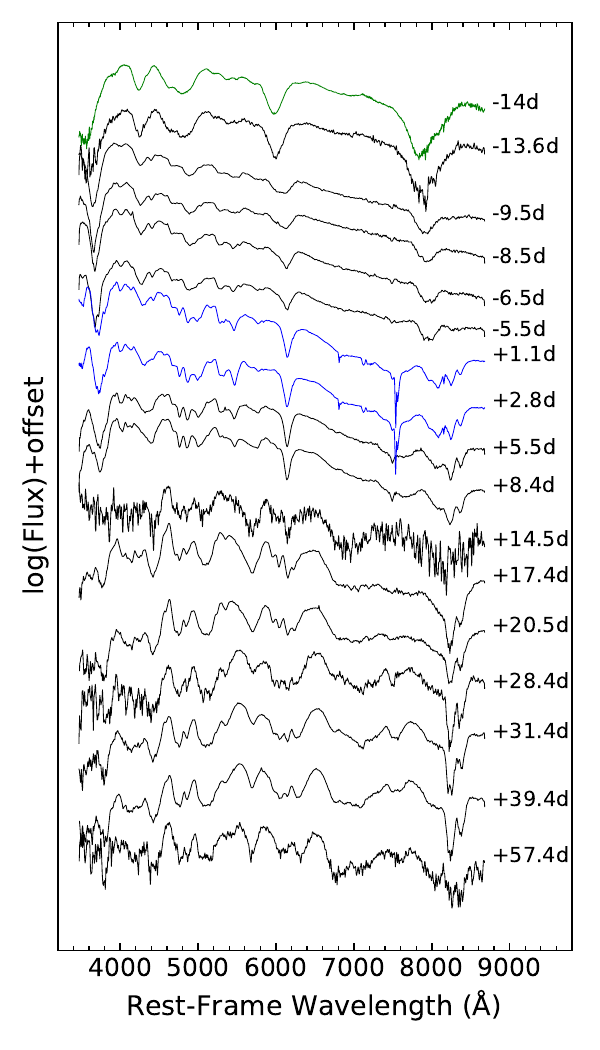}
	\caption{Optical spectral evolution of SN 2021fxy. The spectra have been corrected for the redshift of the host galaxy and telluric lines. The first spectra from TNS marked in green. Two spectra (not corrected for telluric lines) aroud the maximum light marked in blue have been published in \citet{2023MNRAS.522.3481D}. The numbers on the right-hand side mark the epochs of the spectra in days after the $B$-band maximum.  \label{fig:Sp}}
\end{figure}

Fourteen low-resolution spectra of SN 2021fxy collected by LJT, spanning from $t = -14$ d to $t = +57$ d, are presented in Figure \ref{fig:Sp}. 
The corresponding information about the spectra is listed in Table \ref{tab:spec}. 
For further analysis and discussion, a spectrum (marked in green in Figure \ref{fig:Sp}) obtained on March 18, 2021, in SALT with RSS \citep{2015ATel.7119....1F} presented in Transient Name Sever \citep[TNS \footnote{\url{https://www.wis-tns.org/object/2021fxy}};][]{2012PASP..124..668Y} is also plotted. 
Two spectra (marked in blue in Figure \ref{fig:Sp}) published in \citet{2023MNRAS.522.3481D} are included. 
In addition, the continuum flux of these spectra have been corrected with the \textit{BVgri}-band photometry at similar phases.

\section{Photometry Analysis}\label{sec:3}
\subsection{Optical Light Curves} \label{subsec:LC}

\begin{figure}
	\includegraphics[width=\columnwidth]{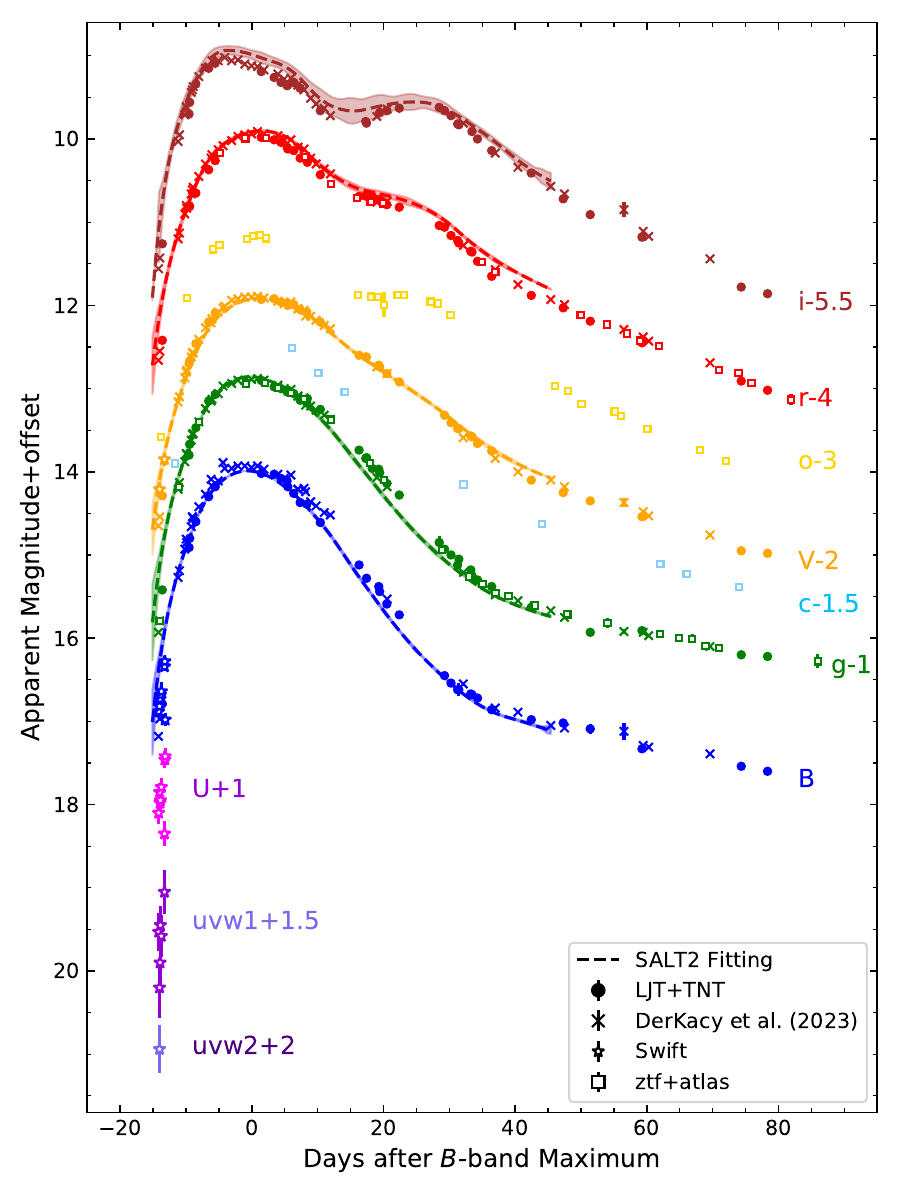}
	\caption{Ultraviolet and optical light curves of SN 2021fxy. The light curves are shifted vertically for better display. The dashed lines are the best fittings of SALT2 model. The regions with different colors are the  1$\sigma$ confidence interval of the best fittings.\label{fig:LC}}
\end{figure}
Figure \ref{fig:LC} displays the optical and UV light curves of SN 2021fxy.
Owing to our sparse temporal coverage near maximum light, we integrate the published photometry from \citet{2023MNRAS.522.3481D} to improve the robustness of our peak magnitude and decline-rate estimates. 
All subsequent analysis is performed on the combined light curves from LJT and \citet{2023MNRAS.522.3481D}.
For comparison, we also plot public survey photometry from ZTF \citep[Zwicky Transient Facility\footnote{\url{https://www.ztf.caltech.edu/}}; $g$ and $r$ bands;][]{2019PASP..131a8002B,2019PASP..131g8001G} and ATLAS \citep[Asteroid Terrestrial-impact Last Alert System\footnote{\url{https://fallingstar.com/home.php}}; $c$ and $o$ bands;][]{2018PASP..130f4505T} in Figure \ref{fig:LC}. These survey data are used for visualization only and are excluded from the quantitative analyses that follow.

We fit the light curves with polynomials to determine the peak magnitudes, dates at the maximum brightness, and decline rates.
The fitting results are listed in Table \ref{tab:lc_param}. 
SN 2021fxy reaches the $B$-band peak magnitude of $B_{\max}$ = 13.98 $\pm$ 0.02 mag on MJD 59305.18 $\pm$ 0.30.
After maximum, it declines with a rate of $\Delta$$m_{15}(B)$ = 0.94 $\pm$ 0.04 mag.

In addition, we fit the optical light curves of SN 2021fxy to obtain some relevant parameters using the SALT2 fitter \citep[version 2.4 with JLA training set;][]{2014A&A...568A..22B}.
The best fitting results are shown in Figure \ref{fig:LC} (the dashed lines and color regions).
The best fitting parameters are listed in Table \ref{tab:S2_param}.
The results give that the $B$-band peak magnitude of SN 2021fxy is $B_{\max}$ = 14.00 $\pm$ 0.02 mag on MJD 59305.38 $\pm$ 0.02, consistent with that of the polynomial fitting.	
We calculate the $\Delta m_{15}(B)$ and distance modulus using parameter $c$ and$x1$  \citep{2018PASP..130f4101V},  yielding $\Delta$$m_{15}(B)=$  1.10 $\pm$ 0.06 mag and a distance modulus of $\mu_{0}=$ 33.0 $\pm$ 0.1 mag. 

We derive our final parameters by averaging the results from the different methods to reduce systematic uncertainty,
obtaining $B_{\max}=  13.99\pm 0.02$ mag, $T(B)_{\max}={\rm MJD} ~59305.3\pm0.25$ and $\Delta$$m_{15}(B)=1.02\pm0.05$  mag.
These are the final light-curve parameters for SN 2021fxy, used for the analyses in the rest of this paper.

The light curve parameters of SN 2021fxy were also given in \citet{2023MNRAS.522.3481D} using SuperNovae in Object Oriented Python (SNooPy). 
They obtained a dereddened $B_{\max}$ = 13.57 $\pm$ 0.01 mag on MJD 59305.12 $\pm$ 0.34 with  $\Delta m_{15}(B)$ = 1.05 $\pm$ 0.06 mag. 
Their results are consistent with our derived parameters for SN 2021fxy, which are only listed here for a simple comparison and not used in the following analyses.


As shown in the light curves of SN 2021fxy, the prominent secondary peaks in the $i$ band and the shoulders in $r$ band are similar to other normal supernovae. It seems that a ``shoulder''  similar to the $r$ band can also be seen in the $o$ band at the same position.

\begin{deluxetable}{ccccc}
	\tablecaption{Light-curve Parameters of SN 2021fxy \label{tab:lc_param}}
	\tablehead{
		\colhead{Band} & \colhead{$\lambda_{\rm eff}$} & \colhead{$t_{\max}$\tablenotemark{$\dagger$}} & \colhead{$m_{\rm peak}$\tablenotemark{$\dagger$}} & \colhead{$\Delta m_{15}$\tablenotemark{$\dagger$}} \\
		\colhead{ } & \colhead{(\AA)} & \colhead{(JD-2459000.5)} & \colhead{(mag)} & \colhead{(mag)} 
	}
	\startdata
	$B$ & 4450 & 305.18(30) & 13.98(03) & 0.94(04) \\
	$V$ & 5500 & 306.1(30) & 13.92(03) & 0.60(05) \\
	$g$ & 4754 & 306.5(30) & 13.88(02) & 0.75(04) \\
	$r$ & 6196 & 306.6(30) & 13.93(03) & 0.73(03) \\
	$i$ & 7690 & 302.0(40) & 14.54(04) & 0.65(05) \\
	\enddata
	\tablenotetext{\dagger}{The numbers in the parentheses are the 1-$\sigma$ uncertainties in units of 0.01.}
\end{deluxetable}

\begin{table}	
	\caption{Parameters of SN 2021fxy given by SALT2 Model \label{tab:S2_param}}
	\begin{tabular}{ccc}
		\hline 
		\hline 
		Parameter & Value \\
		\hline
		DayMax & 59305.38 $\pm$ 0.02 \\
		Color & 0.04 $\pm$ 0.02 \\
		X1 & $-$0.03 $\pm$ 0.02 \\
		RestFrameMag\_0\_B & 14.00 $\pm$ 0.02 mag \\
		RestFrameMag\_0\_V & 13.93 $\pm$ 0.01 mag \\
		\hline
	\end{tabular}
\end{table}

\begin{figure*}
	\includegraphics[width=\textwidth]{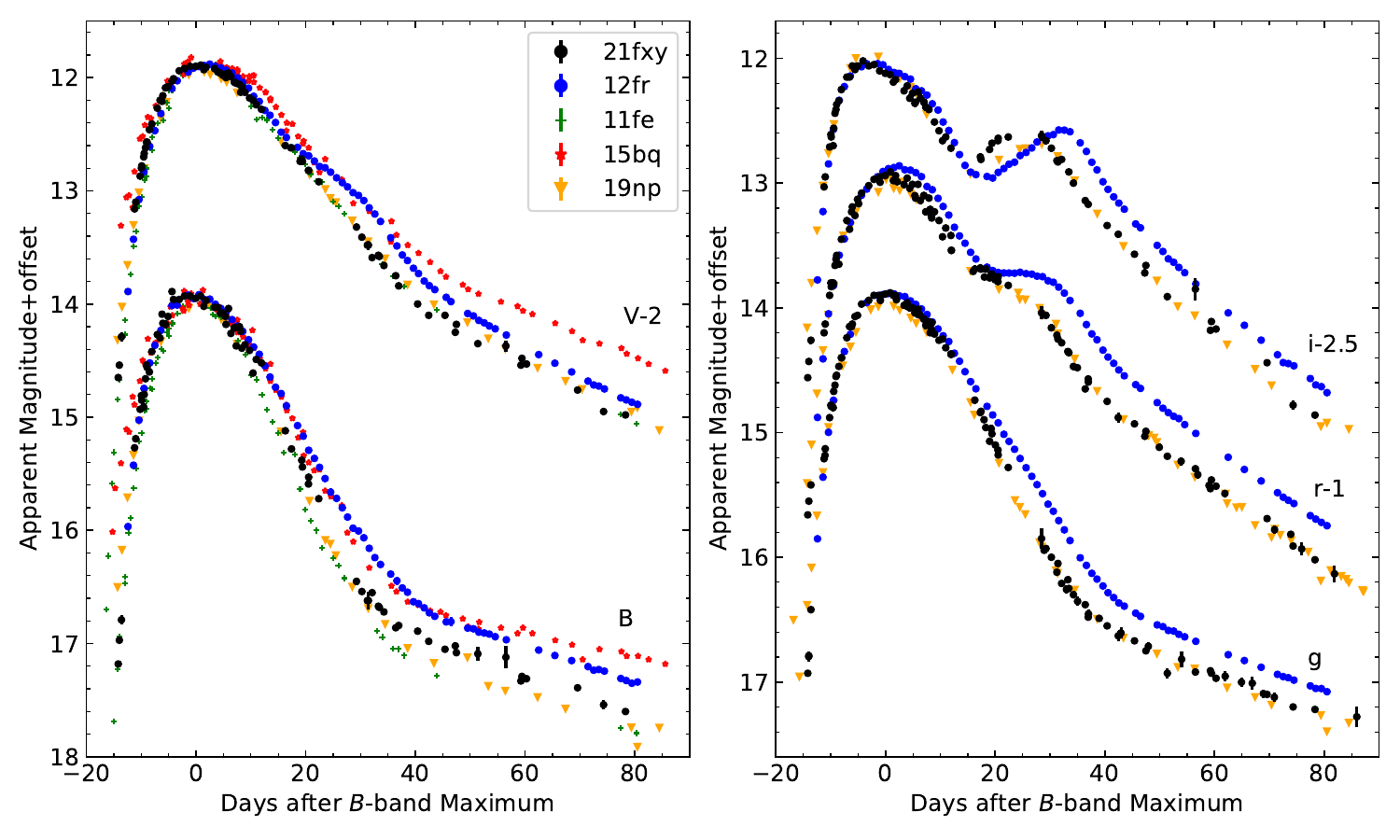}
	\caption{Comparison of the optical light curves of SN 2021fxy to those of other well-observed SNe Ia, including SN 2011fe, SN 2012fr, SN 2015bq, SN 2019np. The light curves of the comparison SNe Ia are	normalized to match the peak magnitudes of SN 2021fxy. \label{fig:LCc}}
\end{figure*}

Figure \ref{fig:LCc} displays the comparison of the light curves of SN 2021fxy with some well-sampled light curves of the SNe Ia, including SN 2015bq \citep[$\Delta$$m_{15}$ = 0.82 mag;][]{2022ApJ...924...35L}, SN Ia 2012fr \citep[$\Delta$$m_{15}$ = 0.85 mag;][]{2014AJ....148....1Z,2018ApJ...859...24C}, SN 2019np \citep[$\Delta$$m_{15}$ = 1.04 mag;][]{2022MNRAS.514.3541S}, and the standard normal-type SN Ia SN 2011fe \citep[$\Delta$$m_{15}$ = 1.08 mag;][]{2012ApJ...753...22B}.

The comparison shows that the light curves of SN 2021fxy resemble those of other normal SNe Ia (e.g. SN 2019np and SN 2011fe) before the maximum.
For instance, the \textit{BV}-band light curves of SN 2021fxy are well matched to those of SN 2011fe at early phases, deviating from those of the luminous SN 2015bq. 
It suggests that there is no early flux excess for SN 2021fxy.
After the maximum, the light curves of SN 2021fxy decline very fast, similar to those of SN 2011fe.
Between $\sim t\approx +30$ d to +80 d, the \textit{B}-band light curve of SN 2021fxy seems to brighter than those of other normal SNe Ia, while the \textit{V}-band light curve is matched well.
There are secondary maximums in \textit{ri} band for SN 2021fxy.
They are weaker and occur earlier than those of SN 2012fr.
Interestingly, the spectra and the spectral evolution of SN 2021fxy and SN 2012fr are similar, although their light curves are totally different after maximum.  
Nevertheless, SN 2021fxy is an SN Ia with normal luminosity according to the light curves.

\subsection{Extinction and Color curves} \label{subsec:cc}
\begin{figure}
	\includegraphics[width=\columnwidth]{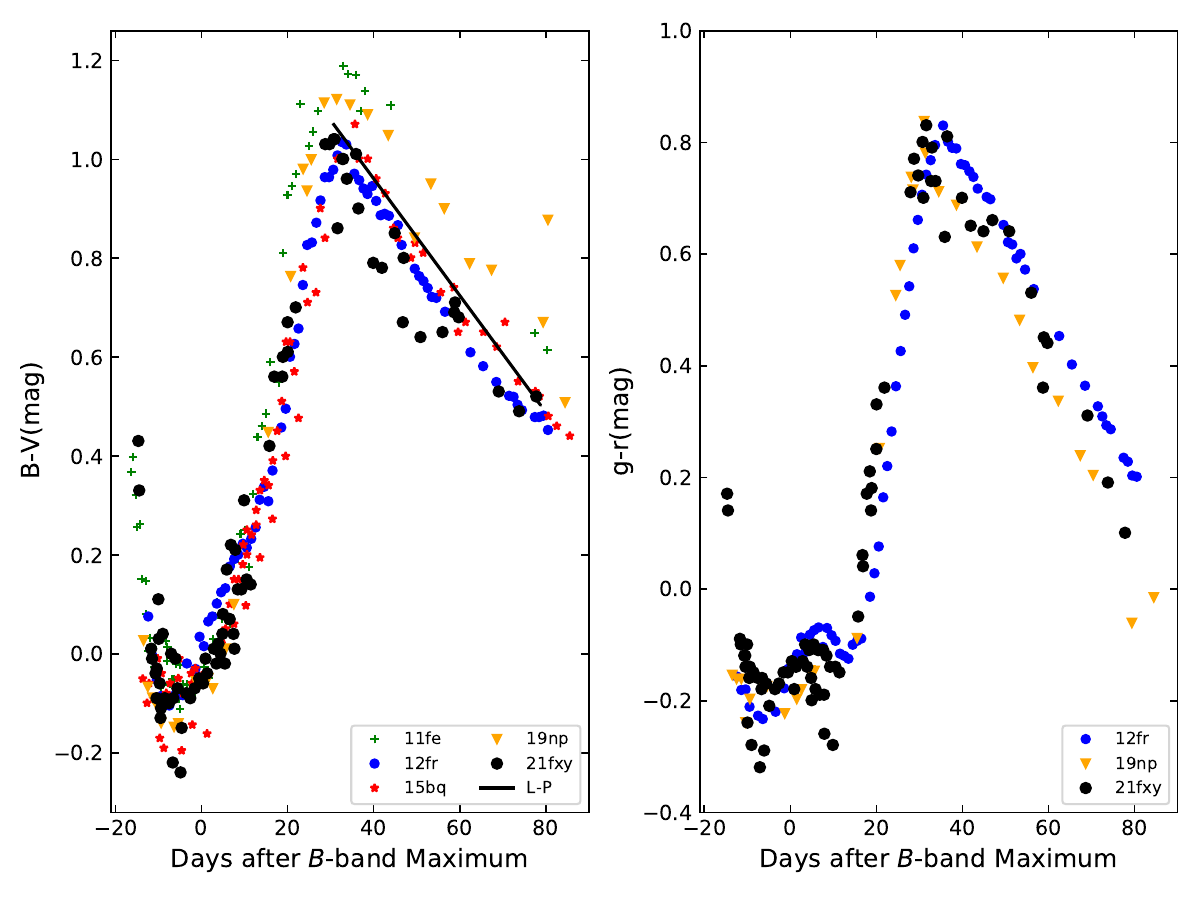}
	\caption{Optical color curves of SN 2021fxy compared with SN 2011fe, SN 2012fr, SN 2015bq, SN 2019np. The black line in the left panel is the L-P relation. See the text for details. \label{fig:CC}}
\end{figure}

The Galactic extinction toward SN 2021fxy is estimated as $A_{V}$(Gal) = 0.261 mag according to the dust map derived by \citep{2011ApJ...737..103S}. Adopting the extinction law with extinction ratio of $R_{V}$ = 3.1, the reddening due to Milky Way is $E(B - V)_{\rm Gal} = 0.084$ mag.


As shown in Figure \ref{fig:CC}, the $B-V$ color of SN 2021fxy at +30 d $ <  t  < $ +90 d seems to deviate from the typical trend of Lira-Phillip (L-P) relation.
The reddening estimation by matching the color curve to the L-P relation may arise a large uncertainty.
To obtain the host reddening, we fit the light curves using the \texttt{color\_model} in SNooPY with the fixed $R_{V}$ of 3.1.
The fitting yields a $E(B - V)_{\rm host} = 0.014\pm 0.06$ mag, consistent with the reddening reported by \citet{2023MNRAS.522.3481D}.
Thus the total reddening for SN 2021fxy is $E(B - V)$ = 0.10 $\pm$ 0.06 mag.
We note that while SNooPy accounts for the correlation between light curve shape and intrinsic color, its color templates are primarily constructed from normal-velocity SNe Ia. As shown by \citet{2019ApJ...882..120W}, high-velocity objects exhibit larger intrinsic color dispersion. If SN 2021fxy possesses strong HVFs, its intrinsic colors may deviate from the SNooPy templates, potentially leading to a slight overestimation of $E(B-V)$ by misattributing intrinsic reddening to dust extinction.

Figure \ref{fig:CC} shows the optical color curves of the same SNe in Figure \ref{fig:LCc}. All color curves are corrected for both Galactic and host-galaxy extinction using the $E(B - V)$ values derived above.

SN 2021fxy and other sources show roughly the same evolution in $B - V$ color. At $t \geq 10$ d, the $B - V$ curve of each supernova gradually turns red until $t \approx +30$ d after the maximum light.
The color curves of normal SNe Ia including that of SN 2021fxy show a similar ``blue-red-blue'' evolution, while that of SN 2015bq shows monotonous reddening at early phase.
Particularly, the $B - V$ color curve of SN 2021fxy exhibits a significant deviation from the L-P relation after $t \approx +30$ d, becoming bluer by $\sim0.2$ mag.
Such a $B - V$ color-curve behavior is suggested to be a characteristic of HV group in the Wang diagram  \citep{2019ApJ...882..120W}.
However, SN 2021fxy is indeed a member of NV group, based on the low velocity of photospheric component of \SiII~\ld6355 (see the analysis in Section \ref{subsec:vel}). 
\citet{2019ApJ...882..120W} suggested that this color-curve behavior may be a light echoes by a cold CSM at a distance of $\sim10^{17}$ cm.

In $g - r$ color,  
SN 2012fr shows a redder color than SN 2021fxy and SN 2019np at about $t\approx  +30$ d. It might be related to the more significant $r$-band shoulder at this phase. 
In addition, the $g - r$ color curves of those objects are well matched, indicating that the reddening estimated for SN 2021fxy is correct.

\subsection{Distance and Luminosity}	
SN 2002dj, a previous supernova in the same host galaxy NGC 5018, yields a distance modulus of $\mu_{0}$ = 32.89 $\pm$ 0.25 mag from SALT2 modeling \citep{2013MNRAS.433.2240G}. A SALT2 fit to SN 2021fxy gives $\mu_{0}$ = 33.0 $\pm$ 0.1 mag. 
We adopt the mean of these two estimates, $\mu_0 = 32.95 \pm 0.19$ mag, for SN 2021fxy and use it throughout this work. This value is consistent with $\mu_0 = 32.87 \pm 0.09$ mag reported by \cite{2023MNRAS.522.3481D}.

With the distance given above, the peak absolute magnitude of SN 2021fxy is then calculated to be $ M_{\max}(B) = -19.36 \pm 0.31 $ mag.
It is close to the peak absolute magnitude of normal SNe Ia ($-19.33\pm0.06$ mag, \citealp{2006ApJ...645..488W}), which is consistent with the result given by the light curve analysis above. With the decline rate of SN 2021fxy ($\Delta m_{15}(B) \sim$ 1.02 mag), one can know that SN 2021fxy is well consistent with the WLR.

To estimate the peak luminosity of SN 2021fxy, we build the pseudo-bolometric luminosity curve by integrating the flux density in the observed $BVri$ bands, with each data point corrected for Milky Way and host-galaxy extinction using the $E(B-V)$ values derived above.
As described in Appendix \ref{sec:append}, the $BV$-band and $gri$-band magnitudes used for the flux integration are in the Vega and AB magnitude system, respectively.
The UV- and NIR-band flux is estimated using the normal flux fractions of SN Ia template.
According to the SALT2 template \citep[][]{2014A&A...568A..22B}, the \textit{U}-band flux is $\sim 15\%$ of the optical band for a normal SN Ia around maximum light.
The UV- and NIR-band flux around maximum can be estimated to be $\sim15\%$ and $\sim5\%$ of the optical flux respectively \citep[e.g.,][]{2009ApJ...697..380W,2012ApJ...749..126W}.
Summing the estimations and \textit{BVri}-band flux, the peak luminosity of SN 2021fxy is thus derived to be $(1.3\pm0.3)\times10^{43}\ \rm{erg\ s^{-1}}$.
Given the possible UV suppression in SN 2021fxy reported by \citet{2023MNRAS.522.3481D}, we note that adopting generic template fractions such as from SALT2 for UV and NIR may bias the peak luminosity.

Fitting the $t^2$ model to the pseudo-bolometric light curve in the rise-time region \citep[$t\leq -10$ d;][]{2012ApJ...745...44G,2015MNRAS.446.3895F}, we obtain the rise time of $t_{\rm r}=16.9\pm0.5$ days for SN 2021fxy.
With the peak luminosity and rise time, a synthesized $^{56}$Ni mass of $M_{\rm ^{56}Ni} = 0.58\pm0.14$ \Msun\,is estimated using the Arnett law \citep[e.g.,][]{1982ApJ...253..785A,2005A&A...431..423S}.

\section{Spectroscopy} \label{sec:4} 
\begin{figure*}
	\includegraphics[width=\textwidth]{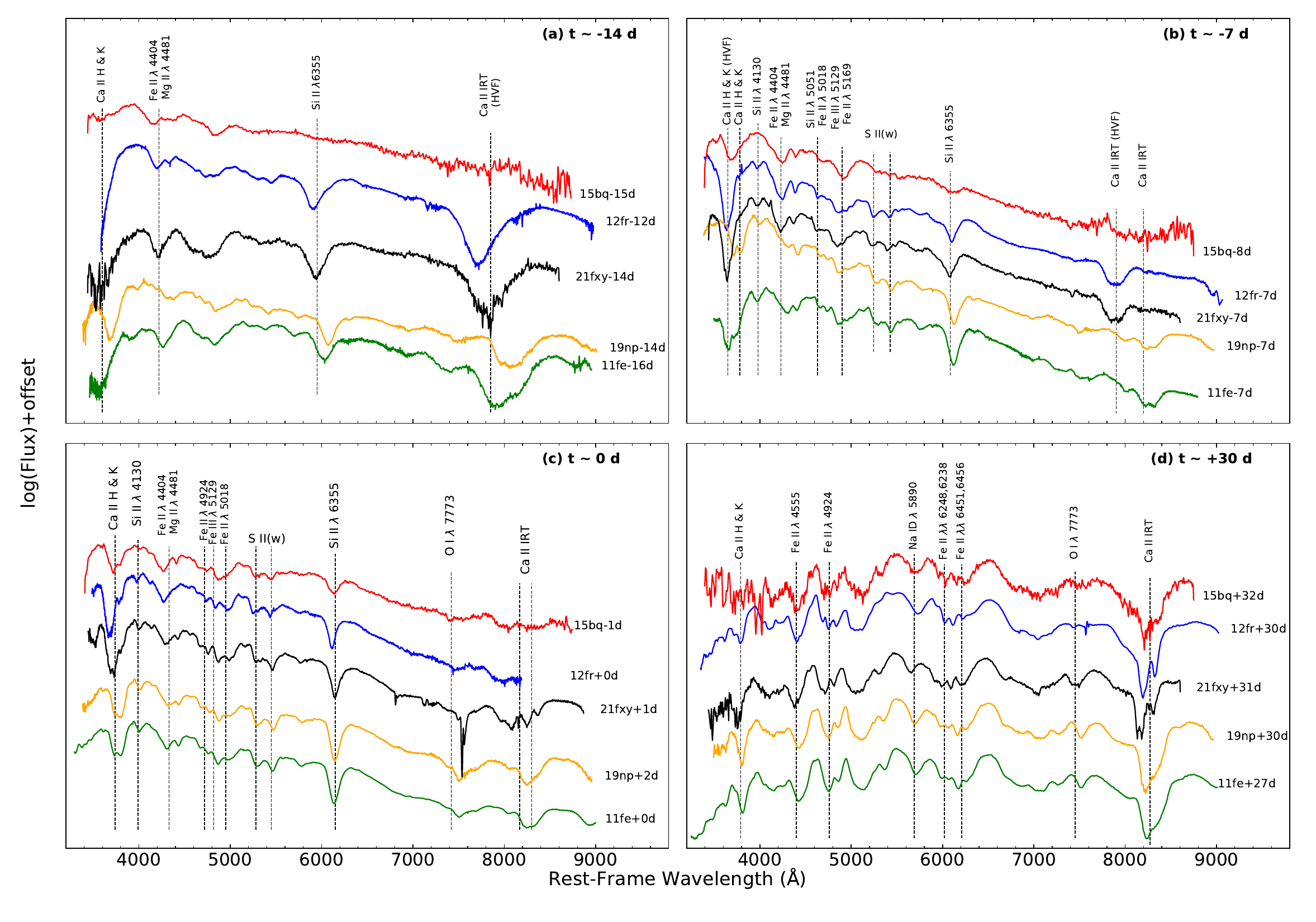}
	\caption{Spectra of SN 2021fxy (black) at $t \sim-$14, $-$7, 0, and +30 days after the \textit{B}-band maximum. The spectra marked in different colors are the comparable-phase spectra of SN 2011fe, SN 2012fr, SN 2015bq, SN 2019np.
		The vertical dash lines represent the wavelengths of the corresponding absorption lines of SN 2021fxy. 
		\label{fig:spec_comp}}
\end{figure*}	

Figure \ref{fig:Sp} shows the spectral evolution of SN 2021fxy.  The early spectra are composed by the absorption of the IMEs, e.g., \ion{Si}{2}, \ion{Ca}{2} H{$\&$}K, ``W''-shaped \ion{S}{2}, and \ion{Ca}{2} near infrared triplet (\CaII~IRT).
These features indicate that SN 2021fxy is basically a normal SN Ia.
In particular, there are prominent HVFs of \SiII~\ld6355 and \CaII~IRT in the early spectra.
Those HVFs are detached from the PVFs and seem to evolve independently. 

\subsection{Temporally Spectral Evolution} \label{sec:spec}
We analyze the temporal evolution and comparison of the spectra at some selected phases of our sample in detail, including SN 2011fe, SN 2012fr, SN 2015bq and SN 2019np (same as the comparison sample of light curves).

The spectra obtained about two weeks before the peak are plotted in Figure \ref{fig:spec_comp}(a). 
The dominant features at this phase are IMEs and iron group elements (IGEs). 
The spectra of SN 2021fxy and SN 2012fr are similar, exhibiting the strong HVFs of \SiII~\ld6355 and \CaII~IRT without PVFs.
The spectra of SN 2019np and SN 2011fe only show the PVFs with lower velocities.
The spectrum of the more luminous SN 2015bq is relatively ``featureless''.

Figure \ref{fig:spec_comp}(b) displays the spectra at $t \approx -7$ d. 
In the spectra of SN 2021fxy, IGE and W-shaped \SII~lines can be read out clearly at this phase.
The \CaII~IRT line still exhibits strong HVF without PVF, while the \SiII~\ld6355 line shows a weak HVF alongside a dominate PVF.
The spectra of SN 2021fxy and SN 2012fr remain closely similar.
No HVFs of \SiII~\ld6355 can be found in the spectra of SN 2019np and 2011fe.
Despite the \SiII~\ld6355 and \CaII~IRT features, the other absorption lines are similar across the spectra of those SNe Ia.

The spectral evolution at around the $B$-band maximum is displayed in Figure \ref{fig:spec_comp}(c). 
The spectrum of SN 2021fxy follows that of a normal SN Ia.
These SNe Ia are generally similar at this phase. However, there are slight differences in the absorption line profile of \ion{Si}{2} $\lambda$6355. 
We calculate the pseudo equivalent widths (pEWs) of \ion{Si}{2} $\lambda$6355 and \ion{Si}{2} $\lambda$5972 near the maximum light for SN 2021fxy. The results are $82.2\pm0.3$ \AA ~and $18\pm5$ \AA ~for \ion{Si}{2} $\lambda$6355 and \ion{Si}{2} $\lambda$5972 respectively.
It suggests that it can be classified into SS subclass in Branch diagram, where SN 2015bq and SN 2012fr are also supposed to be.

The spectra in one month after the $B$-band maximum is displayed in Figure \ref{fig:spec_comp}(d). 
The spectra are dominated by features from IGEs and IMEs, including prominent absorptions from \FeII, \NaI, \CaII~H\&K, and \CaII~IRT. 
The double absorption features of \CaII~IRT are more significant in the spectra of SN 2015bq, SN 2021fxy, and SN 2012fr. 
At this phase, those SNe Ia show a similar spectral evolution.

\subsection{The Ejecta Velocity} \label{subsec:vel}
To measure the ejecta velocities for SN 2021fxy, we fit the absorption profiles in the redshift-corrected spectra with Gaussian functions. For \SII, \CaII~H$\&$K, \SiII~$\lambda$6355, we use double Gaussian functions to fit the lines. The single Gaussian is used in the \SiII~$\lambda6355$ fitting after the \textit{B}-band maximum, since there are only PVFs in the spectra. 
For \CaII~IRT, we use triple Gaussian functions to fit the lines.
We force the function to have three same widths and set the the absorption intensities to follow a proportion of $1:9:5$ ($\lambda8498 : 8542 : 8662$), reducing the free parameters to ensure the parameters are constrained.
Indeed, the proportion of the intensities for the triplet depends on the optical thickness, although the thickness do not significantly affect the fitting results in practice, as discussed in \citet{2013ApJ...770...29C}.
The proportion we adopted in the fitting is based on the assumption of the optically thin regime, obtained using the \textit{gf} values in NIST\footnote{\url{https://www.nist.gov/pml/atomic-spectra-database}}.

The fitting procedures are performed using the Markov chain Monte Carlo (MCMC) method within the \texttt{emcee} package \citep[][]{2013PASP..125..306F}.
We started by selecting the fitting windows of the lines in the spectra manually.
The windows cover the full absorption profiles we interested, including some window-edge data points that are obviously separated from the absorption.
We fit a straight line to those points to measure the ``pseudo-continuum''.
Combining the Gaussian models, the absorption profiles are fitted in the fitting windows.
The velocities are then obtained through the fittings.
The statistical uncertainties of the velocities can be simply obtained from the MCMC samples.
Following \citet{2015MNRAS.446..354P}, we also calculate the uncertainties caused by the manual selection of the fitting windows,  by randomly shifting their ends by approximately ±5 \AA~with respect to the initial positions.
The total uncertainties are finally given by combining the statistical and window-selected uncertainties.

\begin{figure}
	\includegraphics[width=\columnwidth]{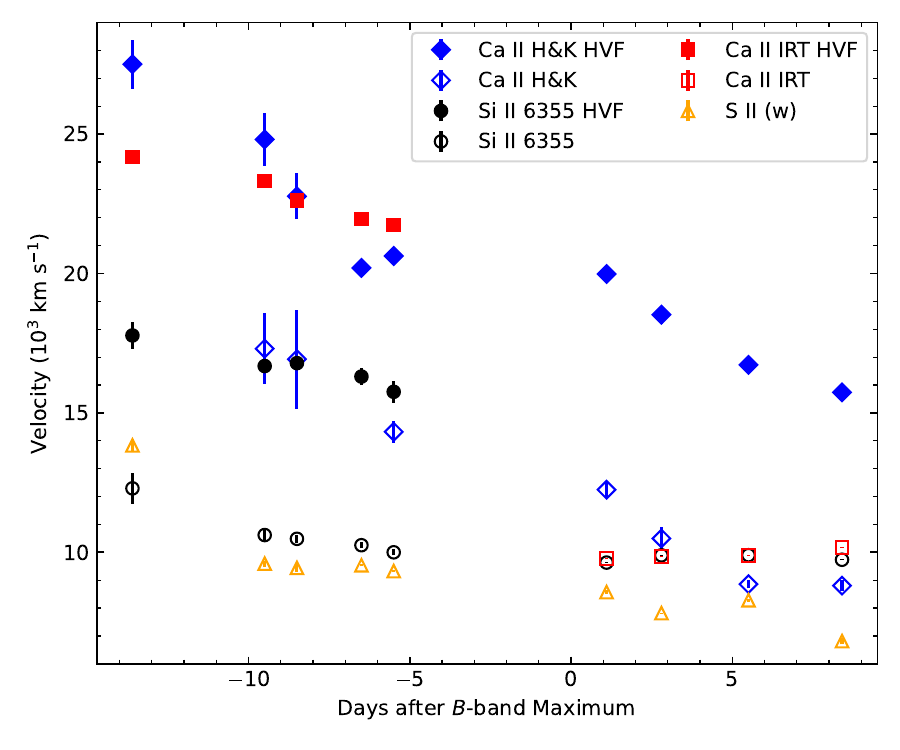}
	\caption{Ejecta velocity evolution of different elements of SN 2021fxy. The solid markers represent the velocities of the HVFs, while the hollow markers represent those of PVFs.  \label{fig:Vel}}
\end{figure}

The velocities of the spectral lines of SN 2021fxy are displayed in Figure \ref{fig:Vel}.  
Note that all velocities are only corrected for the host galaxy redshift.

At $\sim-14$d, the HVFs of the \SiII~\ld6355 and \CaII~IRT exhibit relatively high velocities, reaching $\sim18,000$ and $\sim24,000~\rm{km~s^{-1}}$ respectively.
These velocities subsequently decline at lower rates, compared to those of \CaII~H\&K.

The velocity of the PVF of \SiII~\ld6355 at $-14$ d is $\sim13,000\rm~{km~s^{-1}}$, separated about $5,000\rm~{km~s^{-1}}$ from the HVF.
It then declines gradually and remain a velocity of $\sim10,000~\rm{ km~s^{-1}}$ after the maximum.
Therefore, one can classify SN 2021fxy into NV SNe Ia subclass in Wang diagram \citep{2009ApJ...699L.139W}.
Indeed, the PVF velocities of \ion{Si}{2} $\lambda$6355, \ion{Ca}{2} H{$\&$}K, and \ion{Ca}{2} IRT remain at $ \sim $ 10,000 km $\rm s^{-1}$ for 10 days with very low velocity gradients. It indicates that SN 2021fxy could be classified into the LVG category of SNe Ia in the Benetti diagram \citep{2005ApJ...623.1011B}.

\subsection{Evolution of \SiII~\ld6355 and \CaII~IRT}\label{subsec:HVFs}
\begin{figure*}
\centering
	\includegraphics[width=0.8\textwidth]{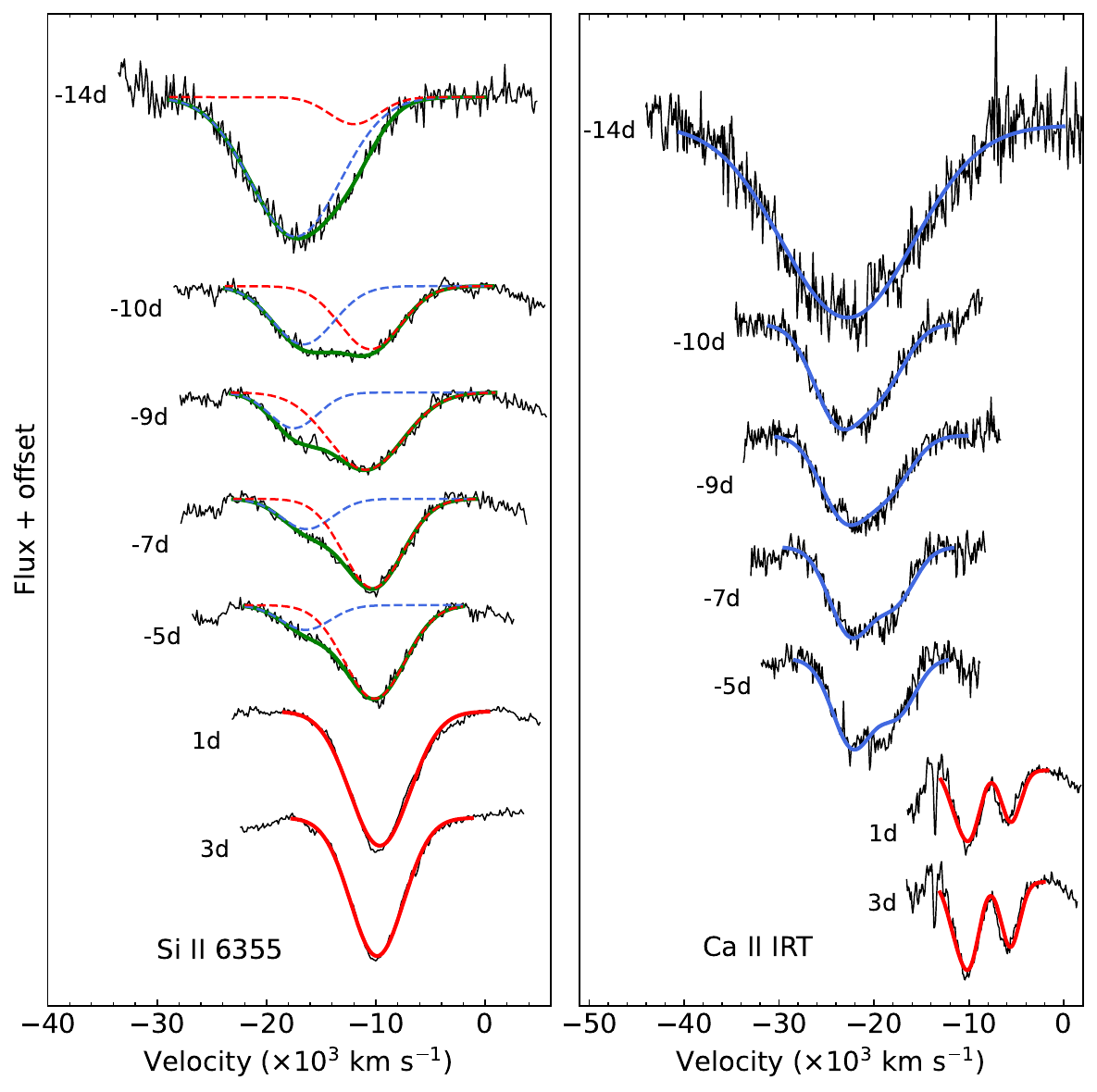}
	\caption{Evolution of HVFs of \ion{Si}{2} $\lambda$6355 (left panel) and  \ion{Ca}{2} IRT (right panel) in the spectra of SN 2021fxy. In the left panel, the blue and red dashed lines represent the Gaussian fits to the detached high-velocity and the photospheric components, respectively. The green dotted line denotes the sum of these two components. In the right panel, the blue and red lines are the best fits to the high-velocity and the photospheric components of \ion{Ca}{2} IRT lines, respectively.\label{fig:HVF}}
\end{figure*}
Figure \ref{fig:HVF} shows the complete evolution of \SiII~\ld6355 and \CaII~IRT, to investigate the HVFs in detail.
The color lines are the best fits to the corresponding absorption profiles, which are described in detail in Section \ref{subsec:vel}.
The figures are plotted in velocity space to show the evolution clearly.
The corresponding velocities of PVFs and HVFs for the lines, obtained using the fitting procedures, are listed in Table \ref{tab:velocity}.
We note that the resolution of the spectra obtained by LJT is $R\approx300$, corresponding to $\Delta v\approx1,000~\rm{km~s^{-1}}$.
The HVFs separated by $\sim 5,000~\rm{km~s^{-1}}$ can be resolved at this spectral resolution.

\begin{deluxetable*}{ccccccccc}
	\tablecaption{Velocities of \SiII~\ld6355 and \CaII~IRT for SN 2021fxy \label{tab:velocity}}
	\tablehead{
		\colhead{Phase} & \multicolumn{4}{c}{\SiII~\ld6355} &  \multicolumn{4}{c}{\CaII~IRT} \\
        \cmidrule(lr){2-5} \cmidrule(lr){6-9}
        \colhead{} & \multicolumn{2}{c}{HVF} &  \multicolumn{2}{c}{PVF} & \multicolumn{2}{c}{HVF} &  \multicolumn{2}{c}{PVF} \\
        \cmidrule(lr){2-3} \cmidrule(lr){4-5} \cmidrule(lr){6-7} \cmidrule(lr){8-9}
		\colhead{ } & \colhead{$v$} & \colhead{pEW} & \colhead{$v$} & \colhead{pEW} & \colhead{$v$} &\colhead{pEW} & \colhead{$v$} & \colhead{pEW} \\
		\colhead{(Days) } & \colhead{($10^3~{\rm km~s^{-1}}$)} &  \colhead{(\AA)} & \colhead{($10^3~{\rm km~s^{-1}}$)} & \colhead{(\AA)} & \colhead{($10^3~{\rm km~s^{-1}}$)} &  \colhead{(\AA)} & \colhead{($10^3~{\rm km~s^{-1}}$)} & \colhead{(\AA)}
	}
	\startdata
    $-13.6$ & $17.8\pm0.5$& $128\pm8$ & $12.3\pm0.6$& $14\pm8$ & $24.2\pm0.1$& $384\pm6$ &  $\cdots$&  $\cdots$\\
    $-9.5$ & $16.7\pm0.2$& $33\pm3$ & $10.6\pm0.2$& $40\pm3$ & $23.32\pm0.05$& $106\pm2$ & $\cdots$&  $\cdots$\\
    $-8.5$ & $16.8\pm0.2$& $25\pm2$ & $10.5\pm0.1$& $48\pm2$ & $22.61\pm0.07$& $91\pm1$ & $\cdots$& $\cdots$ \\
    $-6.5$ & $16.3\pm0.3$& $17\pm2$ & $10.2\pm0.1$& $54\pm2$ & $21.95\pm0.05$& $80\pm2$ & $\cdots$& $\cdots$ \\
    $-5.5$ & $15.7\pm0.4$& $15\pm3$& $10.0\pm0.1$& $51\pm3$ & $21.74\pm0.06$& $82\pm2$ & $\cdots$&   $\cdots$\\
    $1.1$ & $\cdots$& $\cdots$& $9.63\pm0.01$& $82.2\pm0.3$ & $\cdots$ &$\cdots$ &  $9.78\pm0.03$ &  $41\pm1$\\
    $2.8$ & $\cdots$& $\cdots$& $9.90\pm0.01$& $75.6\pm0.5$ & $\cdots$&$\cdots$ &  $9.85\pm0.03$ & $50\pm2$\\
    $5.5$ & $\cdots$& $\cdots$& $9.90\pm0.02$& $73.5\pm0.5$ & $\cdots$&$\cdots$ &  $9.89\pm0.03$ & $60\pm2$\\
    $8.4$ & $\cdots$& $\cdots$& $9.73\pm0.02$& $57.8\pm0.5$ & $\cdots$&$\cdots$ &  $10.18\pm0.03$ & $74\pm2$\\
	\enddata
\end{deluxetable*}

The evolution of \ion{Si}{2} $\lambda$6355 is displayed in the left panel of Figure \ref{fig:HVF}. 
The absorption of \ion{Si}{2} $\lambda$6355 in early spectrum at $t\approx -$14d was dominated by HVF. 
At  $t\approx-10$ d, the HVF and photospheric component appear to have comparable strengths, forming an overall absorption profile with a plateau.
After that, the HVFs of \SiII~\ld6355 quickly decay, while the PVFs dominate the absorption profile. 
Throughout the evolution, the HVFs and PVFs evolve independently, with their velocities decreasing very slowly.

The right panel of Figure \ref{fig:HVF} shows the evolution of \ion{Ca}{2} IRT absorption. 
In the early phase, the HVFs dominate the line profiles, reaching a very high velocity of $\sim25,000~\rm{km~s^{-1}}$.
The substructures of \ion{Ca}{2} $\lambda$8498, $\lambda$8542 and $\lambda$8662 absorptions are cannot be clearly distinguished due to the line blending. 
Before reaching maximum light, the photospheric components are not discernible in the spectrum. 
Around the \textit{B}-band maximum, the PVF of \CaII~IRT is prominent, with a velocity of $10,000~\rm{km~s^{-1}}$.
It gradually strengthens and remains strong for a month after maximum light.
Meanwhile, the velocities of HVFs of \CaII~IRT also decline slowly, while those of the PVFs remain nearly constant after the maximum.


\section{Discussion}\label{sec:5}	
\subsection{Velocity Evolution of the HVFs}\label{subsec:HVFevol}
The ejecta expands freely with the frozen velocity after the explosion, if there is no circumstellar material (CSM).
In homologous expansion, one can derive the ejecta velocity at the photosphere of $v_{\rm ph}\propto t_{\rm exp}^{-2/(n-1)}$ (where $t_{\rm exp}$ is the time after explosion), assuming a density profile of $\rho \propto r^{-n}$ in the steep outer region \citep[e.g.,][]{2010ApJ...708.1025K,2013ApJ...769...67P,2023ApJ...959..132N}.
If the HVFs are caused by the outermost ejecta of the same steep outer region, they may follow similar velocity evolution with the PVFs.

Adopting the typical profile of $n=10$ for SNe Ia \citep[][]{2010ApJ...708.1025K}, the photosphere evolution in the outer region is $v_{\rm ph}\propto t_{\rm exp}^{-0.22}$ \citep[][]{2013ApJ...769...67P,2023ApJ...959..132N}.
We fit the early-phase ($t<-5$ days) PVFs of \SiII~\ld6355 for SN 2021fxy with a power law of $v\propto (t-t_{0})^{-0.22\pm0.02}$, using the MCMC method within \texttt{emcee}. 
The deviation of 0.02 is adopted in the fitting, considering the possible uncertainties from the limited data points or the limitation of our velocity error measurement.
The fitting yields the explosion time of $t_{0}=-17.6\pm0.5$ days after the \textit{B}-band maximum.
This time is $\sim0.7$ days earlier than the rise time of $t_{\rm r}\approx 16.9$ days.
The measured interval of 0.7 days corresponds to the post-explosion dark phase, consistent with that expected for a normal SN Ia.
It indicates that there is no early-excess flux for SN 2021fxy, consistent with the description in Section \ref{subsec:LC}.

\begin{figure}
	\includegraphics[width=\columnwidth]{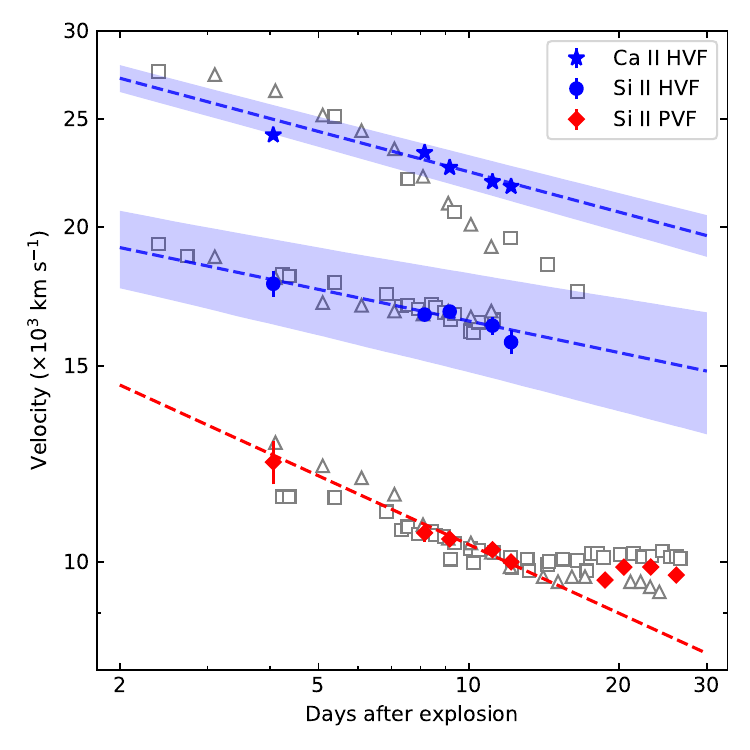}
	\caption{The temporal evolution of the HVFs of SN 2021fxy for \SiII\,\ld6355 and \CaII~IRT. The red dashed line is the best fit to the PVF velocities before $t\approx-5$ days with the fixed power-law index of $-0.22$. The blue dashed lines are the best fits to the HVF velocities of \SiII~\ld6355 and \CaII~IRT, using the power laws with the estimated $t_{\rm exp}$. The blue regions are the $1\sigma$ confidence intervals for the corresponding fits. The gray hollow points are the velocities of  \SiII~\ld6355 HVFs  and PVFs from SN 2012fr (squares) and SN 2009ig (triangles), scaled to match the data of SN 2021fxy. \label{fig:HVF_evol}}
\end{figure}

With the derived $t_{0}$, we fit a power law of $v\propto t_{\rm exp}^{-\beta}$ to the HVF data of \SiII~\ld6355 and \CaII~IRT using MCMC method.
We treat the uncertainty of $t_{0}$ as a Gaussian prior to propagate uncertainty through the fitting process.
The fitting results yield the the HVF evolution of $\beta_{\rm Si} = 0.09\pm0.03$ and $\beta_{\rm Ca} = 0.11\pm0.01$ for \SiII\, and \CaII~respectively, shown in Figure \ref{fig:HVF_evol}.
The data of SN 2009ig \citep{2013ApJ...777...40M} and SN 2012fr \citep[][]{2013ApJ...770...29C,2014AJ....148....1Z} are also plotted in the same figure for comparison.
The velocity evolution of the \SiII~\ld6355 is closely similar across the those objects.
Furthermore, for the HVFs, the evolution shown in SN 2009ig and SN 2012fr is in remarkable agreement with that of SN 2021fxy, falling entirely within the $1\sigma$ confidence interval of the fit described above.
It suggests that $v\propto t_{\rm exp}^{-0.1}$ may be a common velocity evolution for the prominently detached \SiII~\ld6355 HVFs.
This relatively slow evolution deviates significantly from the evolution of the PVFs ($\beta \approx 0.22$) at the same phases, indicating that the HVF ejecta may be a distinct structural component independent of the outer region.
Nevertheless, the evolution of the \CaII~IRT HVFs for SN 2021fxy appears distinct from that in other objects.
This discrepancy may be caused by the measurement bias, as accurately isolating the correct \CaII~IRT lines from their heavily blended absorption profiles in early-phase spectra is particularly challenging.

If the HVF ejecta also follows a power-law density profile of $\rho_{\rm HVF} \propto r^{-n'}$ in a homologous expansion (i.e., $\beta=2/(n'-1)$), a profile of $n'= 22_{-5}^{+10}$ is derived using the velocity evolution of $\beta_{\rm Si} \approx 0.1$ obtained above.
This profile deviates significantly from that suggested for the outer regions of SNe Ia. 
It is even steeper than the profile of $n\approx15$ reported by \citet{2023MNRAS.520..560H} for SN 2019np recently.
This extreme steep profile could not arise from an standard delayed detonation near Chandrasekhar mass \citep[][]{2010ApJ...708.1025K}.
If the HVFs arise from the He shell detonation at the progenitor surface, the double-detonation model may explain this profile naturally. 
However, the profile uncertainty propagated from the other fitted parameters are relatively large.
Any physical derivation based on this profile may unreliable.

In addition, the velocity evolution of HVFs could be attributed to an interaction between the outermost ejecta and the CSM.
In the self-similar solutions of the interaction scenario given by \citet{1982ApJ...258..790C}, the ejecta velocity at photosphere can be roughly described as $v\propto t^{(s-3)/(n-s)}$, with an ejecta density of $\rho \propto r^{-n}$ and a CSM density of $\rho \propto r^{-s}$.
Adopting an ejecta profile of $n\approx11$ and the solar wind CSM profile of $s\approx2$, one can indeed obtain a velocity evolution of $\sim t_{\rm exp}^{-0.09}$ for the HVFs.
However, a typical interaction with a thin CSM of $\leq0.01$ \Msun~may only produce the excess flux lasting one to a few days \citep[e.g., ][]{2021ApJ...923L...8J}, while the HVFs of SN 2021fxy exhibit a coherent evolution that persists for $\sim 10$ days (see Figure \ref{fig:HVF_evol}).
It is unknown whether the thin CSM can effectively affect the explosion dynamics for such a long time.

\citet{2023ApJ...959..132N} reported that the HVFs of SN 2021aefx at very early phases ($t_{\rm exp}<3$ days) follow an evolution similar to that of PVFs ($\sim t_{\rm exp}^{-0.22}$).
They suggested the HVFs may originate from an asymmetric subsonic explosion that produces the split-velocity ejecta in the outermost region with the similar density profile of $\rho \propto r^{-10} $.
However, they also noted that the velocity plateau observed in the HVF evolution at later phases cannot be explained within this framework.
Given the potential measurement biases, it is possible that the reported ``velocity plateau'' in SN 2021aefx is instead a slow evolution comparable to that observed in SN 2021fxy.
Indeed, the assumption of power-law profiles for HVF ejecta may be unnecessary.
The structures isolated from the continually power-law outer region, e.g., the blobs caused by the deflagration fingers \citep[][]{2005ApJ...623L..37M,2013MNRAS.429.1156S}, the He-shell burning ashes \citep[e.g.,][]{2011ApJ...738L...5P,2021ApJ...919..126B}, possibly have different density profiles.
For instance, the slow velocity evolution of the HVFs observed in SN 2021fxy could be caused by the decreasing optical depth of a blob with a Gaussian density profile during its free-expansion phase, with the corresponding absorption feature persisting for about 10 days.
More objects with well-sampled early-phase data, such as SN 2021fxy, SN 2012fr, and SN 2009ig, need to be investigated in detail to confirm the possible HVF evolutionary patterns and their underlying structures.

\subsection{Possible Origin of the Detached HVFs}	
The origins for HVFs typically fall into two broad categories: intrinsic explosion mechanisms and external ejecta-CSM interaction.  
The intrinsic explosion mechanisms include delayed-detonation models that produce IME blobs \citep[e.g.,][]{2013MNRAS.429.1156S} and double-detonation models with He-shell burning ashes \citep[e.g.,][]{2021ApJ...919..126B}.
Conversely, the interaction of ejecta with CSM, like massive clouds, rings, or shells \citep[e.g.,][]{2004ApJ...607..391G,2017MNRAS.467..778M}, can be the possible external mechanism capable of generating the same observational signatures.
Nevertheless, explanations for the HVFs typically involve at least one of the following three mechanisms: abundance enhancements (AE), density enhancements (DE), or ionization effects (IE) \citep[e.g.,][]{2005MNRAS.357..200M,2005ApJ...623L..37M}. 

Many observational studies suggest that the delayed detonations could intrinsically produce the HVFs \citep[e.g.,][]{2020ApJ...897..159P,2023ApJ...959..132N,2025ApJ...986...68Z,2026ApJ...996...10L}.
For instance, given the low excitation potentials for \CaII~($\leq 1.7$ eV), mixing even a small amount of calcium into the outermost ejecta could lead to a local DE during the explosion, resulting in the prominent \CaII~HVFs observed \citep[e.g.,][]{2026ApJ...996...10L}.
However, the strong \SiII~HVFs with higher excitation potentials (8.1 eV) also occur in the spectra of SN 2021fxy simultaneously, yielding a stronger DE or AE process similar to the mixing.
In addition, the uncommon velocity evolution of \SiII~HVFs provides further constrains on their origin, e.g., the HVF ejecta is possibly the structures independent of the outer region, as discussed in Section \ref{subsec:HVFevol}. 
If the IME blobs caused by the deflagration fingers extending into the outer region are produced and account for the HVF ejecta, the \SiII~HVFs could be still involved in the delayed-detonation scenario.

Double-detonation models with thin He shells ($10^{-3}$-$10^{-2}$ \Msun) can produce HVF ejecta \citep[e.g.,][]{2019ApJ...873...84P,2021ApJ...919..126B}, in which IMEs may be synthesized during the detonation of the He shell and thus naturally separated from the underlying progenitor material.
The IME ashes with a possible different density profile could cause a distinct AE/DE in the outermost layers during the explosion, resulting in the uncommon velocity evolution of \SiII~HVFs observed in SN 2021fxy.
Some observational studies suggest HVFs may originate from the double detonation \citep[e.g.,][]{2021ApJ...906...99L}.
However, the early flux excesses in the light curves, which are often associated with He-shell detonations \citep[][]{2019ApJ...873...84P,2021ApJ...923L...8J}, are not detected in SN 2021fxy.
\citet{2021ApJ...919..126B} suggested that the detonations in a thin He shell may produce the light curves without early excesses and synthesize sufficient IMEs to cause their HVFs. 
Similar to the He-shell detonation scenario, He-shell burning on a near-Chandrasekhar-mass WD is also proposed to explain the the prominent \SiII~\ld6355 HVFs \citep[][]{2011ApJ...738L...5P,2018ApJ...863..125K}.

Regardless of the explosion mechanisms themselves, the presence of CSM near the progenitor is the other possible origin for HVFs \citep[e.g.,][]{2004ApJ...607..391G,2017MNRAS.467..778M}.
If the outermost ejecta interacts with CSM, shock compression at the interaction front could produce a local DE.
This additional dense structure may then give rise to the observed HVF absorption.
However, as discussed in Section \ref{subsec:HVFevol}, the CSM interaction may persist for only a few days after explosion, which cannot account for the longer-duration HVF evolution observed.
Although a more massive CSM could extend the interaction timescale, it would in turn predict the presence of strong hydrogen emission lines and possible early excesses --- features that are absent in the observations of SN 2021fxy.
A low-mass CSM with abundant hydrogen may provide sufficient electrons to suppress the excitations of outermost IMEs, forming the broader absorption profile of \CaII~IRT \citep[][]{2005MNRAS.357..200M,2026ApJ...996...10L}.
Considering the temperature of outermost region and the high excitation potentials, strong \SiII~HVFs are unlikely to be produced only through interaction with a low-mass CSM.
The strong \SiII~HVFs may require a more efficient DE or AE process.

The observed mid-UV flux suppression in SN 2021fxy \citep{2023MNRAS.522.3481D} has been interpreted as line blanketing from high-velocity ejecta containing IGEs. 
The IGEs in the outermost ejecta suggests strong mixing process or the presence of distinct outer structures enriched with these elements, which may also contain abundant IMEs. 
Consequently, the HVFs more possibly originate from the explosion mechanism itself, not from CSM interaction. 
The $B-V$ color curve of SN 2021fxy deviating from the LP relation, shown in Section \ref{subsec:cc}, indicates the possible presence of CSM at large distance ($\sim 10^{17}$ cm) from progenitor \citep[][]{2019ApJ...882..120W}.
Such an interaction cannot occur at such a large distance within a few days after explosion and therefore cannot produce the HVFs.

Recently, \citet{2026A&A...707A..57H} studied the HVFs of \SiII~and \CaII~using the spectral modelling method with the radiative-transfer code \texttt{TARDIS}.
They analyzed the spectra of several SNe Ia with significantly detached HVFs, including the data of SN 2021fxy from \citet{2023MNRAS.522.3481D}.
The \SiII~HVFs were successfully reproduced by adding a Gaussian enhancement in the same density profile.
It indicates that the detached HVFs can arise from intrinsic ejecta density and/or abundance enhancements without requiring CSM interaction.
The observed behavior of SN 2021fxy shown in this paper is consistent with those results from this theoretical work, suggesting the intrinsic origin of HVFs.
However, the observed HVFs may arise from a more complex mechanism, rather than being governed by a single factor such as density.
Density, composition, and IE may be coupled, leading to degeneracy in reproducing the observed HVFs \citep{2026A&A...707A..57H}.

In summary, the detached HVFs of SN 2021fxy are possibly caused by density structures separated from the outer region with different density profiles.
The observed velocity evolution of HVFs is consistent with this scenario.
Both delayed-detonation and double-detonation models may be capable of producing the density structures to explain the HVFs. 
However, based solely on the velocity evolution from the current observations, the specific physical model responsible for SN 2021fxy cannot be definitively identified without dedicated modeling.
The CSM interaction scenario is unlikely to be the origin of the HVFs.


\subsection{Spectroscopic Diversities} \label{diversity}
\citet{2014AJ....148....1Z} proposed the photometric properties of the SS SN 2012fr might be similar to that of normal SNe Ia.
Similarly, SN 2021fxy is a member of SS groups, who exhibits normal-like properties in light curves and spectra.
It is also an NV SN Ia that shows some color and/or spectroscopic properties presented in HV events.
It indicates that the diversity of SNe Ia may vary continuously across the parameter space.
Such possible continuum in the subclasses reflects that the diversity in SNe Ia could arise from the differences of underlying physical processes during the explosion, e.g. the temperature, the asymmetric ejecta, the environment, etc., rather than from the totally different explosion mechanism.

Indeed, many statistical studies support the proposition that the diversity within the SN Ia population stems from underlying continuous distributions.
\citet{2024ApJ...969...80C} and \citet{2026arXiv260202677S} consistently showed that SN Ia light curve properties may form continuous distributions, based on light-curve modeling and subclass analysis for the ZTF and the Carnegie Supernova Project (CSP) data.
Similarly, \citet{2025A&A...694A...9B} found a continuous spectral distribution from luminous (91T-like) to faint (91bg-like) SNe Ia in a volume-limited ZTF sample.
Some rare objects with the multi-subclass properties are believed to be the evidence for the existence of the continuous distributions.
For instance, SN 1986G and 2012ij are proposed to bridging the gap between normal and 91bg-like group, reported by \citet{2016MNRAS.463.1891A} and \citet{2022ApJ...927..142L} respectively.
It indicates that the gaps between different subclasses can be filled eventually with more observation on such objects, making the observational properties of SNe Ia continuous distributions.

One may suspect that the spectra of those objects like SN 2021fxy/SN 2012fr could be affected by the presence of HVFs, the most significant features different from the other SNe Ia in our analyses, leading to a unreliable classification.
However, although the HVF ejecta may cause the mid-UV suppression \citep[e.g.,][]{2023MNRAS.522.3481D}, there is no evidence shown that the HVFs can cause changes in spectra at maximum light.
\citet{2026MNRAS.546f2281H} found that \SiII~PVFs around maximum  light may not be significantly affected by the occurrence of HVFs.
Whether there are other reasons biasing the classifications for those objects is unknown.

\section{Conclusion}\label{sec:6}
We present the optical observations of SN 2021fxy, covering the phase from $-$14 to $+$78 days relative to its  $ B $-band maximum light. 
These observations show that the photometric properties of SN 2021fxy are broadly consistent with those of normal SNe Ia, with the decline rate of $\Delta m_{15}(B) = 1.02\pm 0.05$ mag and the absolute peak magnitude of $M_{\rm max}(B) = -19.36\pm0.31$ mag.
The corresponding peak luminosity is $(1.3\pm0.3)\times 10^{43}~\rm{erg~s^{-1}}$, yielding a synthesized $^{56}$Ni mass of $0.58\pm0.14$ \Msun. 
In particular, the $B-V$ color curve of SN 2021fxy exhibits a blue offset of $\sim0.2$ mag relative to the L-P relation after $t=+30$ d, while its earlier evolution aligns with that of normal SNe Ia.

The spectra of SN 2021fxy at early phases show strong IME HVFs.
The velocity of the \CaII~IRT HVF at $\sim-14$ days reaches $\sim2,5000~\rm{km~s^{-1}}$.
The \SiII~HVFs are detached from the PVFs by $\sim5,000~\rm{km~s^{-1}}$, lasting until $t\approx-5$ days.
Interestingly, the velocity evolution of the \SiII~and \CaII~IRT HVFs seems to follow relatively shallow power-law declines of $\sim t_{\rm exp}^{-0.1}$.
It significantly deviates from the power-law index of $\sim0.2$ for the PVFs, assuming an $n=10$ density profile in the homologous expansion scenario.
Such velocity evolution indicates that there may be density structures that are independent of the outermost region and have distinct density profiles. 
The observed HVFs could originate from those structures in the explosion.
These structures could be the blobs in delayed detonations or the He-shell ashes in double detonations.
However, the physical origin of the HVFs cannot be constrained by the velocity evolution analysis alone.
More early-phase observations of SNe Ia with \SiII~\ld6355 HVFs and detailed physical modelings are needed to constrain the possible HVF origin and ultimately reveal the underlying explosion mechanism.   

In addition, SN 2021fxy shows some color and/or spectroscopic properties associated with different subclasses.
It suggests that the population diversity in SNe Ia likely arising from continuous variations in key explosion parameters within a unified physical mechanism.
It is unknown whether the presence of HVFs can affect the individual properties and cause the diversity in a subclass.
The observations of the larger sample with \SiII~\ld6355 HVFs like SN 2021fxy could give more details on the diversity in the subclasses, contributing to improving the accuracy of cosmological distance measurements.

	

\section*{Acknowledgments}
	
We thank the anonymous referee for the constructive comments. This work is supported by the National Key R\&D Program of China with grant 2021YFA1600404, the B-type Strategic Priority Program of the Chinese Academy of Sciences (Grant No. XDB1160202), the National Natural Science Foundation of China (NSFC grants 12173082, 12333008, 12225304 and 12225304), the Yunnan Fundamental Research Projects (YFRP; grants 202501AV070012, 202401BC070007, 202201AT070069, and 202501AS070005), the Top-notch Young Talents Program of Yunnan Province, the Light of West China Program provided by the Chinese Academy of Sciences (CAS),  the CAS Project for Young Scientists in Basic Research (YSBR-148), and the International Centre of Supernovae (ICESUN), Yunnan Key Laboratory of Supernova Research (No. 202505AV340004). X.-F.W. is supported by the NSFC (grants 12288102, 12033003, and 11633002) and the Tencent Xplorer Prize. 

We acknowledge the support of the staff of the LJT and TNT. Funding for the LJT has been provided by the CAS and the People’s Government of Yunnan Province. The LJT is jointly operated and administrated by YNAO and Center for Astronomical Mega-Science, CAS.

\bibliography{21fxy}{}
\bibliographystyle{aasjournal}
	
	
	
\appendix
\section{Photometry and spectroscopy data of SN 2021fxy} \label{sec:append}
\setcounter{table}{0} 
\renewcommand{\thetable}{\Alph{section}\arabic{table}}
	
We present the photometric data and the spectroscopic information of SN 2021fxy in this section. Table \ref{tab:stand} gives the photometric standards in Johnson \textit{BV} and Sloan \textit{gri} bands used for the reduction of LJT and TNT in Section \ref{sec:2}.
The photometric data of \textit{gri} bands in AB magnitude system are obtained from Pan-STARRS\footnote{\url{https://catalogs.mast.stsci.edu/panstarrs/}}.
The \textit{BV} magnitudes are obtained by transforming the \textit{gr}-band to \textit{BV}-band magnitudes using the magnitude relations given by \citet{1996AJ....111.1748F}.
Eventually, the \textit{BV}-band (in Vega system) and \textit{gri}-band (in AB system) magnitudes for SN 2021fxy are obtained through differential photometry.
The optical photometry in \textit{BVgri} bands obtained by LJT and TNT is listed in Table \ref{tab:Opti_Pho}.
The \textit{Swift} UVOT photometry plotted in Figure \ref{fig:LC} is shown in Table \ref{tab:UVOT}.
The journal of the spectroscopic observations shown in Figure \ref{fig:Sp} is listed in Table \ref{tab:spec}.	

	\begin{deluxetable}{cccccc}[h]
		\tablecaption{Local Photometric Standards in the \textit{BVgri} bands \label{tab:stand}}
		\tablehead{
			\colhead{Star} & \colhead{$B$ (mag)} & \colhead{$V$ (mag)} & \colhead{$g$ (mag)} & \colhead{$r$ (mag)} &\colhead{$i$ (mag)}
		}
		\startdata
		1 & 16.68(0.04) & 15.94(0.04) & 16.26(0.03) & 15.73(0.03) & 15.48(0.03) \\
		2 & 15.56(0.03) & 14.95(0.03) & 15.20(0.02) & 14.80(0.02) & 14.64(0.02) \\
		3 & 15.03(0.04) & 14.59(0.04) & 14.75(0.03) & 14.51(0.03) & 14.44(0.02) \\
		4 & 16.39(0.04) & 15.82(0.04) & 16.05(0.03) & 15.68(0.03) & 15.54(0.02) \\
		\enddata	
		\tablecomments{See Figure \ref{fig:img} for the finder chart of SN 2021fxy and the comparison stars. The values in parentheses are the corresponding uncertainties (1$\sigma$).}
	\end{deluxetable}
	
	\startlongtable
	\begin{deluxetable*}{cccccccc}
		\tablecaption{Optical Photometry of SN 2021fxy obtained by LJT and TNT \label{tab:Opti_Pho}}
		\tablehead{
			\colhead{MJD} & \colhead{Epoch\tablenotemark{a}} & \colhead{$B$ (mag)} & \colhead{$V$ (mag)} & \colhead{$g$ (mag)} & \colhead{$r$ (mag)} & \colhead{$i$ (mag)} & \colhead{Telescope}
		}
		\startdata
		59291.70 	&	-13.6 	&	16.79(0.04)	&	16.29(0.04)	&	16.42(0.02)	&	16.42(0.02)	&	16.76(0.03)	&	LJT	\\
		59295.77 	&	-9.5 	&	14.91(0.03)	&	14.70(0.03)	&	14.80(0.03)	&	14.80(0.03)	&	15.20(0.03)	&	TNT	\\
		59295.86 	&	-9.4 	&	14.80(0.01)	&	14.67(0.01)	&	14.67(0.01)	&	14.81(0.01)	&	15.06(0.01)	&	LJT	\\
		59296.80 	&	-8.5 	&	14.60(0.03)	&	14.46(0.02)	&	14.47(0.02)	&	14.65(0.01)	&	14.84(0.01)	&	LJT	\\
		59298.74 	&	-6.6 	&	14.30(0.03)	&	14.20(0.02)	&	14.15(0.02)	&	14.37(0.01)	&	14.65(0.01)	&	LJT	\\
		59299.73 	&	-5.6 	&	14.18(0.04)	&	14.09(0.03)	&	14.07(0.02)	&	14.26(0.02)	&	14.59(0.02)	&	LJT	\\
		59306.73 	&	1.4 	&	14.02(0.01)	&	13.93(0.01)	&	13.90(0.01)	&	14.05(0.01)	&	14.69(0.01)	&	LJT	\\
		59308.69 	&	3.4 	&	14.03(0.03)	&	13.92(0.03)	&	13.98(0.03)	&	14.02(0.03)	&	14.76(0.03)	&	TNT	\\
		59309.73 	&	4.4 	&	14.08(0.03)	&	13.96(0.03)	&	14.00(0.03)	&	14.24(0.03)	&	14.96(0.03)	&	TNT	\\
		59310.66 	&	5.4 	&	14.11(0.03)	&	13.97(0.03)	&	14.04(0.03)	&	14.10(0.03)	&	14.86(0.03)	&	TNT	\\
		59310.77 	&	5.5 	&	14.18(0.01)	&	14.00(0.01)	&	14.02(0.01)	&	14.18(0.01)	&	14.85(0.01)	&	LJT	\\
		59311.67 	&	6.4 	&	14.26(0.03)	&	13.99(0.03)	&	14.06(0.03)	&	14.14(0.03)	&	14.82(0.03)	&	TNT	\\
		59312.64 	&	7.3 	&	14.38(0.03)	&	14.05(0.03)	&	14.14(0.03)	&	14.23(0.03)	&	14.89(0.03)	&	TNT	\\
		59313.65 	&	8.4 	&	14.41(0.03)	&	14.08(0.03)	&	14.17(0.03)	&	14.26(0.03)	&	$\cdots$	&	TNT	\\
		59313.73 	&	8.4 	&	$\cdots$	&	$\cdots$	&	14.12(0.08)	&	14.28(0.01)	&	14.05(0.03)	&	LJT	\\
		59315.70 	&	10.4 	&	14.61(0.01)	&	14.20(0.01)	&	14.25(0.01)	&	14.43(0.01)	&	15.16(0.01)	&	LJT	\\
		59321.59 	&	16.3 	&	14.97(0.03)	&	14.68(0.03)	&	14.74(0.03)	&	14.72(0.03)	&	$\cdots$	&	TNT	\\
		59322.60 	&	17.3 	&	$\cdots$	&	14.61(0.03)	&	14.84(0.03)	&	14.78(0.03)	&	15.30(0.03)	&	TNT	\\
		59322.72 	&	17.4 	&	15.26(0.01)	&	14.62(0.01)	&	14.83(0.01)	&	14.79(0.01)	&	15.31(0.01)	&	LJT	\\
		59323.58 	&	18.3 	&	$\cdots$	&	$\cdots$	&	14.96(0.03)	&	14.69(0.03)	&	$\cdots$	&	TNT	\\
		59324.58 	&	19.3 	&	15.35(0.03)	&	14.72(0.03)	&	14.97(0.03)	&	14.76(0.03)	&	$\cdots$	&	TNT	\\
		59324.72 	&	19.4 	&	15.44(0.02)	&	14.74(0.01)	&	15.01(0.01)	&	14.78(0.01)	&	15.17(0.01)	&	LJT	\\
		59325.81 	&	20.5 	&	15.59(0.02)	&	14.86(0.01)	&	15.14(0.01)	&	14.81(0.01)	&	15.18(0.02)	&	LJT	\\
		59327.69 	&	22.4 	&	15.76(0.03)	&	14.92(0.01)	&	15.28(0.02)	&	14.87(0.01)	&	15.19(0.02)	&	LJT	\\
		59333.76 	&	28.5 	&	$\cdots$	&	$\cdots$	&	15.85(0.08)	&	15.14(0.05)	&	15.14(0.04)	&	LJT	\\
		59334.57 	&	29.3 	&	16.59(0.03)	&	15.32(0.03)	&	15.93(0.03)	&	15.06(0.03)	&	15.16(0.03)	&	TNT	\\
		59335.55 	&	30.3 	&	16.61(0.03)	&	15.41(0.03)	&	16.00(0.03)	&	15.16(0.03)	&	15.22(0.03)	&	TNT	\\
		59336.56 	&	31.3 	&	16.62(0.03)	&	15.49(0.03)	&	16.12(0.03)	&	15.22(0.03)	&	15.32(0.03)	&	TNT	\\
		59336.75 	&	31.4 	&	16.62(0.08)	&	15.58(0.04)	&	16.05(0.02)	&	15.31(0.01)	&	15.33(0.01)	&	LJT	\\
		59338.56 	&	33.3 	&	16.67(0.03)	&	15.64(0.03)	&	16.18(0.03)	&	15.35(0.03)	&	$\cdots$	&	TNT	\\
		59338.74 	&	33.4 	&	16.68(0.03)	&	15.69(0.02)	&	16.25(0.02)	&	15.46(0.02)	&	15.41(0.02)	&	LJT	\\
		59339.57 	&	34.3 	&	16.72(0.03)	&	15.66(0.03)	&	16.30(0.03)	&	15.47(0.03)	&	15.52(0.03)	&	TNT	\\
		59341.73 	&	36.4 	&	16.73(0.03)	&	15.75(0.02)	&	16.38(0.01)	&	15.65(0.01)	&	15.64(0.01)	&	LJT	\\
		59347.74 	&	42.4 	&	16.87(0.03)	&	16.24(0.02)	&	16.63(0.04)	&	15.95(0.04)	&	15.93(0.02)	&	LJT	\\
		59352.59 	&	47.3 	&	17.27(0.03)	&	16.29(0.03)	&	$\cdots$	&	16.03(0.03)	&	16.38(0.03)	&	TNT	\\
		59356.70 	&	51.4 	&	17.09(0.06)	&	16.35(0.03)	&	16.93(0.04)	&	16.19(0.02)	&	16.44(0.03)	&	LJT	\\
		59364.55 	&	59.3 	&	17.33(0.02)	&	16.54(0.01)	&	16.91(0.01)	&	16.45(0.01)	&	16.68(0.02)	&	LJT	\\
		59379.63 	&	74.3 	&	17.54(0.04)	&	16.97(0.03)	&	17.20(0.02)	&	17.01(0.02)	&	17.37(0.04)	&	LJT	\\
		59383.63 	&	78.3 	&	17.60(0.03)	&	16.95(0.02)	&	17.18(0.02)	&	17.02(0.02)	&	17.36(0.02)	&	LJT	\\
		\enddata
		\tablecomments{The values in parentheses are the corresponding uncertainties (1$\sigma$); MJD = JD - 2400000.5.}
		\tablenotetext{a}{Referring to the peak of $B$ band on April 01 2021, MJD 59305.3.}
	\end{deluxetable*}
	
	\begin{deluxetable*}{ccccccc}
		\tablecaption{$Swift$ UVOT Photometry of SN 2021fxy \label{tab:UVOT}}
		\tablehead{
			\colhead{MJD} & \colhead{Epoch\tablenotemark{a}} & \colhead{$uvw2$ (mag)}  & \colhead{$uvw1$ (mag)} & \colhead{$U$ (mag)} & \colhead{$B$ (mag)} & \colhead{$V$ (mag)} 
		}
		\startdata
		59291.25 	&	-14.05 	&	$\cdots$	&	18.53(0.23)	&	17.60(0.13)	&	16.88(0.07)	&	$\cdots$	\\
		59291.38 	&	-13.92 	&	19.44(0.29)	&	19.20(0.36)	&	17.35(0.11)	&	16.82(0.07)	&	16.21(0.09)	\\
		59291.46 	&	-13.84 	&	$\cdots$	&	18.90(0.30)	&	17.48(0.12)	&	16.73(0.08)	&	$\cdots$	\\
		59291.51 	&	-13.79 	&	$\cdots$	&	18.45(0.23)	&	17.45(0.12)	&	16.68(0.07)	&	$\cdots$	\\
		59291.66 	&	-13.64 	&	$\cdots$	&	18.58(0.25)	&	17.29(0.11)	&	16.64(0.11)	&	$\cdots$	\\
		59292.13 	&	-13.17 	&	$\cdots$	&	18.05(0.26)	&	17.85(0.15)	&	16.34(0.06)	&	15.85(0.08)	\\
		59292.18 	&	-13.12 	&	$\cdots$	&	$\cdots$	&	16.96(0.10)	&	16.28(0.06)	&	$\cdots$	\\
		59292.24 	&	-13.06 	&	$\cdots$	&	$\cdots$	&	16.92(0.10)	&	16.98(0.08)	&	$\cdots$	\\
		\enddata
		\tablecomments{The values in parentheses are the corresponding uncertainties (1$\sigma$); MJD = JD - 2400000.5.}
		\tablenotetext{a}{Referring to the peak of $B$ band on April 01 2021, MJD 59305.3.}
	\end{deluxetable*}
	
	\begin{deluxetable*}{ccccccc}
		\tablecaption{Spectroscopic observation journal of SN 2021fxy \label{tab:spec}}
		\tablehead{
			\colhead{MJD} & \colhead{Epoch\tablenotemark{a}} & \colhead{Res.} & \colhead{range} & \colhead{Exp. Time} & \colhead{Airmass} & \colhead{Telescope}\\
			\colhead{ } & \colhead{days} & \colhead{(\AA~$\rm{pixel}^{-1}$)} & \colhead{(\AA)} & \colhead{(s)} & \colhead{ } & \colhead{(+Instrument)}
		}
		\startdata
		59291.68 	&	-13.6 	&	25	&	3200-9800	&	1350	&	1.85	&	LJT YFSOC	\\
		59295.83 	&	-9.5 	&	25	&	3200-9801	&	1350	&	1.64	&	LJT YFSOC	\\
		59296.78 	&	-8.5 	&	25	&	3200-9802	&	1200	&	1.46	&	LJT YFSOC	\\
		59298.72 	&	-6.6 	&	25	&	3200-9803	&	1350	&	1.49	&	LJT YFSOC	\\
		59299.71 	&	-5.6 	&	25	&	3200-9804	&	1350	&	1.52	&	LJT YFSOC	\\
		59310.75 	&	5.5 	&	25	&	3200-9805	&	900	&	1.47	&	LJT YFSOC	\\
		59313.71 	&	8.4 	&	25	&	3200-9806	&	900	&	1.45	&	LJT YFSOC	\\
		59319.82 	&	14.5 	&	25	&	3200-9807	&	1200	&	2.21	&	LJT YFSOC	\\
		59322.70 	&	17.4 	&	25	&	3200-9808	&	900	&	1.45	&	LJT YFSOC	\\
		59325.79 	&	20.5 	&	25	&	3200-9809	&	900	&	2.01	&	LJT YFSOC	\\
		59333.74 	&	28.4 	&	25	&	3200-9810	&	1200	&	1.7	&	LJT YFSOC	\\
		59336.73 	&	31.4 	&	25	&	3200-9811	&	1350	&	1.66	&	LJT YFSOC	\\
		59344.70 	&	39.4 	&	25	&	3200-9812	&	1500	&	1.62	&	LJT YFSOC	\\
		59362.70 	&	57.4 	&	25	&	3200-9813	&	1500	&	2.21	&	LJT YFSOC	\\
		\enddata
		\tablenotetext{a}{Referring to the peak of $B$ band on April 01 2021, MJD 59305.3.}
	\end{deluxetable*}

\end{document}